\def\sollum {\hbox{$L_{\odot}$}}
\def\lesssim{\mathrel{\hbox{\rlap{\hbox{\lower4pt\hbox{$\sim$}}}\hbox{$<$}}}}
\def\gtrsim{\mathrel{\hbox{\rlap{\hbox{\lower4pt\hbox{$\sim$}}}\hbox{$>$}}}}
\def\arcsec {\hbox{$^{\prime\prime}$}}
\def\deg{\ifmmode^\circ\else$^\circ$\fi}
\documentstyle[11pt,aaspp]{article}
\received{1 August 1996}
\accepted{10 December 1996}
\begin{document}

\title{A Survey for H$_2$O Megamasers in Active Galactic Nuclei -- II. A
Comparison of Detected and Undetected Galaxies}

\author{J. A. Braatz}
\affil{Department of Astronomy, University of Maryland, College Park, MD 20742}
\author{A. S. Wilson}
\affil{Department of Astronomy, University of Maryland, College Park, MD 20742
\\and\\Space Telescope Science Institute, 3700 San Martin Drive, Baltimore, MD 21218}
\and
\author{C. Henkel}
\affil{Max-Planck-Institut f\"{u}r Radioastronomie, Auf dem H\"{u}gel 69,
D-53121 Bonn, Germany}

\begin{abstract}

A survey for H$_2$O megamaser emission from 354 active galaxies has resulted
in the detection of 10 new sources, making 16 known altogether.  The galaxies
surveyed include a distance-limited sample (covering Seyferts and LINERs
with recession velocities $<$ 7000 km s$^{-1}$) and a magnitude-limited sample
(covering Seyferts and LINERs with m$_B$ $\leq$ 14.5).  In order to determine
whether the H$_2$O-detected galaxies are ``typical'' AGN or have special
properties which facilitate the production of powerful masers, we have
accumulated a database of physical, morphological, and spectroscopic
properties of the observed galaxies.  The most significant finding is that
H$_2$O megamasers are detected only in Seyfert 2 and LINER galaxies, not 
Seyfert 1s.
This lack of detection in Seyfert 1s indicates that either they do not have
molecular gas in their nuclei with physical conditions appropriate to produce
1.3 cm H$_2$O masers or the masers are beamed away from Earth, presumably in
the plane of the putative molecular torus which hides the Seyfert 1 nucleus in
Seyfert 2s.  LINERs are detected at a similar rate to Seyfert 2s, constituting
a strong argument that at least some nuclear LINERs are AGN rather than
starbursts, since starbursts have not been detected as H$_2$O megamasers.
We preferentially detect H$_2$O emission from the nearer galaxies and from
those which are apparently brighter at mid- and far-infrared and centimeter
radio wavelengths.  There is also a possible trend for the H$_2$O-detected 
galaxies to be more intrinsically luminous in nuclear 6 cm radio emission
than the undetected ones, though these data are incomplete.
We find evidence that Seyfert 2s with very high (N$_H$ $>$ 10$^{24}$ cm$^{-2}$)
X-ray absorbing columns of gas are more often detected as H$_2$O maser emitters
than Seyfert 2s with lower columns.
It may be that the probability of detecting H$_2$O maser emission in 
Seyfert galaxies increases with increasing column of cool gas to the nucleus, 
from Seyfert 1s through NLXGs to Seyfert 2s.
\end{abstract}


\section{Introduction}

Sub-milliarcsecond VLBI mapping of the H$_2$O maser emission in NGC
4258 shows that the masers originate in a thin disk or annulus with inner
radius 0.13 pc and outer radius 0.28 pc (Miyoshi et al. 1995).  The
edge-on disk is in Keplerian rotation about the (presumed) nuclear black hole
and central engine, and it has
a significant warp.  Parts of the disk are thus
illuminated by the weak, obscured nuclear X-ray source
reported by Makishima et al. (1994).  Heating of molecular gas by X-ray
absorption has been shown to give rise to an H$_2$O-enhanced
layer in which collisional pumping can account for the observed
maser luminosity in NGC 4258 (Neufeld \& Maloney, 1995).

It is not clear whether the geometry of the nuclear gas in NGC 4258 is unique
or common among AGN.  If many AGN have similar nuclear disks which support
powerful H$_2$O masers, then the small number of galaxies with detected H$_2$O
megamasers would be just those for which the nuclear disk is viewed close to
edge-on.  In this case, the isotropic properties of the H$_2$O-detected and
undetected galaxies would be indistinguishable.  On the other hand, 
it may be that some special geometrical or physical conditions, which are
present only in a subset of AGN, are required to produce powerful H$_2$O masers.
The combination of a nuclear hard X-ray source and a thin, warped molecular
disk, as seen in NGC 4258, is one possibility.  
Another is a nuclear hard X-ray source surrounded by a thick torus, with
the maser emission originating from the heated inner edge of torus, as
may be the case in NGC 1068 (Greenhill et al. 1996).  We might then
expect to find other observable manifestations of these special conditions 
in the galaxies detected as H$_2$O megamasers.

Braatz, Wilson \& Henkel (1996, hereafter [Paper I])
presented the results of a large (354 galaxies) survey of
active galaxies (Seyferts, LINERs and radio galaxies) for emission from H$_2$O
in the 6$_{16}$ -- 5$_{23}$ 1.3 cm line.  Ten new H$_2$O 
megamasers\footnote{In Paper I and the present paper, we adopt the
arbitrary definition of an H$_2$O megamaser as a galaxy emitting an isotropic
luminosity of $>$ 20 $\sollum$ in the 6$_{16}$ -- 5$_{23}$ line.} 
were discovered, bringing the total number of such galaxies known to 16.  The
present paper describes a study based on this survey, with the goal of
determining whether the galaxies which are detected in H$_2$O emission are
similar to the general AGN population or represent a special subset. 
To this end, a
database of photometric, morphological and spectroscopic properties of all
active galaxies observed for H$_2$O emission has been compiled.
The galaxies in the database include the 354 galaxies observed in Paper I,
the five previously known H$_2$O megamasers (NGC 1068, NGC 3079, NGC 4258,
NGC 4945 and the Circinus galaxy) and five galaxies observed by
others (M31 and M81 by Huchtmeier, Eckart \& Zensus 1988; 
M51 (NGC 5194) and Arp 220 by Henkel, Wouterloot \& Bally 1986; and NGC 5195 by
Claussen \& Lo 1986), for a total of 364.
Table 1 lists the most important 
entries in the database.  The primary sources of data are: the Huchra 
(1993) and 
V\'eron-Cetty \& V\'eron (1991) catalogs; the Third Reference Catalog
of Bright Galaxies (RC3) (de Vaucouleurs et al.  1991); the NASA Extragalactic
Database (NED); and the IRAS Faint Source Catalog (Moshir et al. 1989).

Section 2 describes the galaxy samples and statistical tests used. 
Section 3 compares the total galaxy properties of the H$_2$O-detected 
and undetected galaxies, while Section 4 is devoted to a similar comparison
for the nuclear properties.  Results and conclusions are summarized in
Section 5.

\section{The Galaxy Samples and Statistical Tests}

For the purpose of statistical analyses, we will consider several subsets of
the list of observed AGN.  The first, which will be referred to as the
``distance-limited sample'', consists of all observed Seyferts and
LINERs with $cz \leq$ 7000 km s$^{-1}$.  The second (``sensitivity-limited
sample'') includes all objects in the distance-limited sample which have
been observed with an absolute sensitivity to 1.3 cm H$_2$O emission of better
than 2 $\sollum$ (km s$^{-1}$)$^{-1}$.  This sample is useful because the
absolute sensitivity varies greatly
from galaxy to galaxy.  The limit chosen is empirical, but ensures that all
galaxies with a detected narrow-line megamaser are included in the
sensitivity-limited sample.  The third sample (``magnitude-limited sample'')
includes all objects
observed from Huchra's catalog with an apparent B magnitude brighter
than or equal to 14.5.  In presenting the statistical results, these samples
are also occasionally divided according to activity class, i.e. Seyfert 1,
Seyfert 2 and LINER.  A summary of the number of galaxies in each of the
various samples is given in Table 2.

In the following sections the galaxies detected in H$_2$O 1.3 cm
emission will be compared to the undetected
galaxies graphically, and statistical tests will be applied to give
a quantitative assessment of the significance of any differences found.
All the histograms have a similar layout, as described in the caption to
Figure 2.  The number of galaxies plotted in each histogram
can be found in Table 3.

In evaluating the plots and results of the statistical tests, it is helpful to
bear in mind the following simplified situations.  If all active galaxies
had similar H$_2$O maser luminosities, then we would tend to detect the
{\it nearer} ones.  If the ratio of the maser luminosity to the luminosities
in other wavebands is similar from galaxy to galaxy (i.e. the galaxies have 
similar spectral shapes, but widely differing intrinsic luminosities), then
we would tend to detect the {\it apparently brighter} (in other wavebands)
ones.  In a dataset with a limited number of detections, it may be difficult
to distinguish between these possibilities, since the nearer galaxies tend to
be the apparently brighter.  We will also search for correlations between
{\it absolute} maser and other waveband luminosities, especially for properties
which may be physically associated with the H$_2$O maser production.
Examples of such properties might be 1.3 cm radio continuum 
luminosity (a source of ``seed'' photons) and nuclear hard X-ray luminosity
(the probable source of heating of the nuclear disk).  
Our hope is to identify which, if
any, intrinsic properties are crucial for H$_2$O maser production in AGN.

\subsection{Description of the Statistical Tests}

In comparing the physical properties of the detected and undetected galaxies,
we will apply several general, nonparametric tests.
When an observation results in no detection but has an upper (or lower) limit,
the resulting data point is referred to as a ``censored'' value.
A measured or calculated data point is an
``uncensored'' value.  For example, the inclinations of the galactic disks are
all uncensored, whereas the IRAS flux at 12 $\mu m$ (which has only an
upper limit for many galaxies) is partially censored.
The Kolmogorov-Smirnov (K-S) test is used to compare distributions which
contain only uncensored data.  The K-S test applies to unbinned
data values, so it has a higher ``resolution'' than the binned data plotted
in the histograms.
In comparing censored data, the techniques of ``survival analysis''
(Feigelson \& Nelson, 1985) have been applied via the ASURV software package.
The five statistical tests from ASURV which are used to compare the
distributions are: Gehan's generalized Wilcoxon test with permutation
variance, Gehan's generalized Wilcoxon test with hypergeometric variance,
the logrank test, the Peto-Peto test, and the Peto-Prentice test.  These
tests are generalizations of standard techniques based on
ranking the data values and comparing the distributions of the ranks.
They differ from one another primarily in the way that censored data points are
weighted in the computation of the test statistic.  All of the weightings are
valid, and the prevailing wisdom says to
consider all results and use caution when the reported statistical 
significances are vastly different.  The Peto-Prentice test has been
shown to be particularly effective when comparing samples with dissimilar
sizes, as we have here for the H$_2$O-detected and undetected galaxies.
The tests from the ASURV package are valid for
uncensored data as well as censored, though for completely uncensored data
the Peto-Peto test reduces to the logrank test and the Peto-Prentice
method reduces to Gehan's test.

The result of each statistical test is a significance value which describes the
probability that one parent population would produce the two samples.
In Table 3, these probabilities are listed for each parameter, mostly for the
distance-limited sample.  A probability
of 0.05 or smaller is often taken to be indicative of a meaningful
difference in the distributions.  A probability of greater than 0.10
indicates there is little evidence that the parent distributions
are different.  Probability values of less than about 0.01
are not well defined for nonparametric tests involving small samples.

\subsection{Distribution in Space}

Figure 1 shows the location in Galactic coordinates for each source
observed.  The data
are plotted in an Aitoff projection so that equal areas on the plot represent
equal areas on the sky.  The zone of avoidance through the Galactic plane is
evident.  Fewer than 6\% of all the observed galaxies are found between
Galactic latitudes --15\deg and 15\deg, an area which covers more than a quarter
of the total sky.  The distribution of observed galaxies is not dominated
by individual clusters -- for example, only 10 galaxies from the Virgo cluster
are included.

A histogram of the recession velocity of observed galaxies is shown in
Figure 2.  The galaxies detected in H$_2$O emission tend towards smaller
recession velocities (i.e. they are closer) than undetected galaxies in
the distance- and magnitude-limited samples.  The probability that the 
distributions of recession velocities of the H$_2$O-detected and undetected
galaxies in the distance-limited sample are different is 91.4\% from the
K-S test, and $\geq$ 99\% from
the other tests (Table 3).  Thus there are certainly more galaxies with
H$_2$O emission than we have found, but they are simply too distant to detect
at our sensitivity.  As might be expected, the distribution of detected
galaxies does follow that of the sensitivity-limited sample.

\subsection{Completeness}

For galaxies within a few hundred Mpc (z $\lesssim$~0.1), evolutionary
and cosmological
effects are minimal, so it is reasonable to assume that Seyfert and LINER 
galaxies within this region are uniformly distributed in space.
Given this presumed uniformity,
it is clear that AGN catalogs are incomplete to the distance and magnitude
limits chosen for our observations, and this incompleteness is reflected in
our samples (any cosmological evolution with the number of active galaxies 
increasing towards larger distance strengthens this conclusion).
Furthermore, a number of known AGNs were
not observed because of telescope scheduling limitations.  In addition to
the 5 galaxies mentioned in Paper I which have either $cz$ $<$
7000 km s$^{-1}$
or m$_B <$ 14, we did not observe 20 galaxies with 14.0 $<$ m$_B <$ 14.5
from the Huchra catalog.  Because we {\it have} observed 20 galaxies from this
last range of magnitudes, we
choose to use the limiting magnitude of 14.5 rather than 14 in our
magnitude-limited sample.

A common test for sample completeness is the $V/V_{max}$ test.  This procedure
was devised by Schmidt (1968) as a test for uniform spatial distribution in
quasar samples.  If it is known beforehand that a sample of objects
is uniformly distributed in space, then the $V/V_{max}$ test can be applied to
check for completeness.  The procedure works as follows:  for each source in
the sample, a sphere, $S$, is defined with a radius equal to the distance
from the Earth to the source.  The volume of $S$ is given by
$V = {4 \over 3} \pi r^3$, where $r$ is the distance to the source.
A second sphere, $S_{max}$, is defined with a radius $r_{max}$ equal to the
maximum distance the source could have and still be a member of the sample.
For example, in a sample selected by the single criterion of being closer
than the distance corresponding to a recessional velocity of 7000 km s$^{-1}$,
the radius of $S_{max}$ would be
$r_{max}$ = 7000 km s$^{-1}$ / H$_0$.  For a sample selected only by having
magnitude brighter than m$_B$ = 14.5, the radius would be
$r^{\prime}_{max} = r \times 10^{(14.5-m_B)/5}$, where $m_B$ is the apparent
magnitude of the galaxy and $r$ its distance.
The values of $V/V_{max}$ for a uniformly distributed
sample of galaxies are expected to have a uniform distribution between 0 and
1, and the mean value $<V/V_{max}>$ = 0.5.
Figure 3 shows the value of $<V/V_{max}>$ as a function of recession velocity
for the distance-limited sample, assuming that
$V_{max}$ is determined only by the recession velocity limit of 7000
km s$^{-1}$.  Figure 4 shows $<V/V_{max}>$ as
a function of magnitude for the magnitude-limited sample, assuming that
$V_{max}$ is determined only by the magnitude limit of m$_B$ = 14.5.
The real situation is clearly more complex, with the degree of completeness
of the two samples being limited by both magnitude and distance, which are
not independent quantities, and other parameters discussed below.
It is, nevertheless, clear that both samples are incomplete, with
$<V/V_{max}>$ = 0.25 $\pm$ 0.02 for the distance-limited sample and
$<V/V_{max}>$ = 0.27 $\pm$ 0.02 for the magnitude-limited sample.
The drop-off from the expected galaxy counts at larger distances
is also evident in Figure 2, in which the number of galaxies
in each velocity bin would grow quadratically with increasing distance for a
uniform and complete sample.  Figures 3 and 4 also show the values of
$<V/V_{max}>$ for each AGN class individually.  The degree of incompleteness
is similar for Seyfert 1
and 2 galaxies, but the LINERs are not as well represented at large distances
or at faint magnitudes.  LINER nuclei have weaker emission lines
than Seyferts, and the high quality, starlight-subtracted
spectra required to distinguish this nuclear emission from that of the
surrounding starlight are only available for nearby galaxies.

The completeness of the samples is primarily determined by the methods and
sensitivities used in surveys for AGN. 
Other factors which contribute to incompleteness include (a)
the increasing difficulty of detecting point-like Seyfert nuclei with increasing
distance (starlight contamination increases with distance for a fixed
aperture size), (b) obscuration in and near the
Galactic plane (zone of avoidance), and (c) the difficulty of detecting 
Seyfert nuclei in highly inclined galaxies because of obscuration
by the galactic disk.  The last effect is well known
(e.g. Keel 1980) and is illustrated for our samples in
Figure 5, where the observed distribution of galactic
inclinations for the distance-limited sample is compared to that for a
group of randomly oriented thin disks.  The distribution of thin disks is
normalized such that the number of disks with inclinations $i <$ 45\deg\ 
is equal to the number of galaxies in the distance-limited sample with
$i <$ 45\deg.  Disregarding incompleteness of the galaxies
in the subset of the distance-limited sample
with $i <$ 45\deg , we find that there are $\sim$~110 galaxies
which are omitted from the distance-limited sample because of internal
obscuration by the galactic disk.  The figure shows an apparent
underrepresentation of
galaxies in the 10\deg\ -- 30\deg\ range and an overabundance of galaxies
in the 30\deg\ -- 40\deg\ range, which may indicate a systematic error in the
measurement of axial ratios, perhaps influenced by the presence of bars or
disturbed systems.

\section{Properties of the Galaxy}

\subsection{B Magnitude}

Figures 6 and 7 show the distributions of apparent and absolute total
(i.e. galaxy plus nucleus) B magnitudes
for the H$_2$O-detected and undetected galaxies.  The integrated m$_B$ 
magnitudes are primarily taken from RC3 and are not corrected for galactic 
extinction.  The detected galaxies tend to be brighter than the undetected
ones in the distance- or 
magnitude-limited samples (with $\sim$~90\% confidence -- see Table 3).
However, there is no difference between the distributions of the
detected galaxies and those undetected in the sensitivity-limited sample.
The difference in apparent brightness between the detected and undetected
galaxies in the
distance- and magnitude-limited samples is therefore best understood as
a sensitivity issue,
as discussed previously in connection with the recession velocities of the
samples.  When the Seyfert 1 galaxies (which are all undetected in H$_2$O
maser emission -- see Section 4.1) are left out of the 
calculation, the difference in the detected and undetected galaxies'
apparent brightness becomes slightly more significant (Table 3).

The absolute magnitudes (M$_B$) of the galaxies were calculated using
``face-on'' apparent magnitudes corrected for internal and Galactic extinction
(from RC3).  The distribution of detected galaxies is not unusual
compared to the undetected ones (Figure 7), although the results given
in Table 3 show a weak tendency for the maser-detected galaxies to be of
lower luminosity.  These calculations (but not the figure) 
include 51 galaxies from the distance-limited sample which 
do not have magnitude corrections available for internal and Galactic 
extinction, but do have upper limits to M$_B$ determined from uncorrected
apparent magnitudes.  The absence of maser detections in Seyfert 1
galaxies does not account for the weak tendency of H$_2$O-detected galaxies to
be intrinsically fainter
than undetected ones.  Table 3 shows the test results comparing detected
galaxies with only undetected Seyfert 2's and LINERs, and the 
marginal difference in the population persists at about the same
confidence level ($\sim$~88\%) as when all undetected galaxies (including
Seyfert 1's) are included.

\subsection{Morphological Type}

The distribution of Hubble stage index is shown in Figure 8.
The index, $T$, as described in RC3, gives a numerical assignment to
the large-scale galactic morphology, from elliptical 
galaxies on the left ($T$ = --5 or --4), to lenticulars
(--3 $\leq$ $T$ $\leq$ --1), and early- to late-type spirals
(0 $\leq$ $T$ $\leq$ 7) on the right.  The values
$T$ = 8 and $T$ = 9 are used to describe galaxies with only a hint of spiral
structure, $T$ = 90 represents irregular galaxies, and $T$ = 99 is reserved for
peculiar galaxies.  The distribution of $T$ among the detected
galaxies is indistinguishable from that of the undetected ones (Table 3).

\subsection{Inclination}

The inclination $i$ of each galactic disk was
calculated from the axial ratio $b/a$ according to
$ i = sin^{-1}\sqrt{[1 - (b/a)^2]/0.96}$ (e.g. Whittle, 1992).  During earlier
stages of this survey, there was a trend for the detected megamasers to prefer
highly inclined host galaxies (Braatz, Wilson \& Henkel, 1995).  A histogram
of inclination angles for the current sample is shown in Figure 9.
The addition of the recently discovered
masers in IC 1481 and NGC 5347, which have modest inclinations,
decreases the probability that the masers come from a more highly inclined 
parent distribution than the undetected galaxies.
The probability that the detected galaxies have higher inclinations than the
undetected ones is in the range 80\% -- 94 \%, depending on which test is used
(Table 3).  Despite the marginal character of this difference,
the substantial fraction of masers in inclined
galaxies (7 of the 14 galaxies with detected megamasers and meaningful
inclinations have $i >$ 67\deg) merits some discussion.

Maiolino \& Rieke (1995, hereafter MR95) have analyzed a magnitude-limited
sample of Seyfert galaxies to study, among other things, the relation of
Seyfert type to disk inclination.  MR95 give special
consideration to intermediate Seyfert classifications
(i.e. 1.2, 1.5, 1.8, 1.9).  They provide evidence that Seyferts with
the intermediate classifications 1.8 or 1.9 favor highly inclined galaxies,
and they suggest that these nuclei may be type 1 Seyferts which are obscured by 
a 100 pc-scale torus in the galactic disk rather than the nuclear torus,
which is probably randomly oriented with respect to the galactic disk
(Ulvestad \& Wilson 1984b; Wilson \& Tsvetanov 1994).
Indeed, three galaxies with detected H$_2$O megamasers
are included in MR95's sample as type 1.9 Seyferts (NGC 2639, NGC 4258,
and NGC 5506) and five other megamasers are included as type 2 Seyferts
(in this paper we classify NGC 2639 and NGC 4258 as LINERs, not Seyferts).
Because classifications into the intermediate types are not available or
are contradictory for many of the galaxies in our sample, we have opted to
list only the basic classification (i.e. type 1, 2, or LINER).
The breakdown of inclinations by Seyfert types
for the distance-limited sample is shown in Figure 10.  Seyfert 1 galaxies are
mildly underrepresented at high inclinations: ${10} \over {43}$ = 23\% of
Seyfert 1s with meaningful inclination values have $i$ $>$ 70\deg, compared to
${33} \over {105}$ = 31\% of Seyfert 2s and ${20} \over {54}$ = 37\% of LINERs.
To determine the impact of this effect on the statistical results, we have
also compared the inclinations of the
H$_2$O-detected galaxies to only Seyfert 2 and LINER undetected
galaxies.  The results are essentially the same as when the undetected Seyfert 1
galaxies are included (Table 3).

If the total gain in extragalactic H$_2$O maser sources were {\it dominated} by
amplification in gas associated with the 100 pc-scale torus of MR95 or other
molecular clouds in the galactic disk, we might expect the total or peak
observed maser
power to be correlated with the galaxy inclination.  Figure 11 shows the
isotropic total maser luminosity of each detected source plotted as a function
of the galactic inclination.  The maser luminosities are taken from
Table 2 of Paper I and are plotted with a nominal
uncertainty of 25\%.  Measured luminosities are subject to calibration
uncertainties, intrinsic variability, and especially the unknown beaming
factor which corrects the inferred ``isotropic luminosity''
to the true luminosity.  Clearly the plot should be interpreted
cautiously; regardless, it shows no trends.
The peak luminosities, obtained from the peak fluxes given
in Table 2 of Paper I, are plotted against the inclination in Figure 12, with 
nominal 50\% uncertainties.  Again, no correlation is evident.

A compelling argument against maser amplification in the disk of the galaxy
has been that the masers are very compact and associated only with the nucleus
(e.g. Claussen \& Lo 1986; Haschick et al.  1990).
If maser amplification occurred in the disk of the host galaxy,
both extended and compact nuclear radio continuum emission could be amplified,
so the observed maser emission would tend to trace
the nuclear 1.3 cm radio continuum structure.  Recently, Gallimore
et al. (1996) have discovered weak H$_2$O maser components in NGC 1068 located
$\sim$~0.3\arcsec\ (20 pc) from the ``true'' nucleus, and coincident with the
brightest 1.3 cm radio continuum peak.  It seems unlikely that these masers
receive significant amplification from disk gas since NGC 1068 is
fairly face-on ($i \simeq$ 29\deg).  Gallimore et al. (1996) suggest that the
nonnuclear masers
are associated with a shock front which excites the ambient molecular gas.

In conclusion, if the possible trend for detected H$_2$O megamasers to occur
preferentially in highly inclined galaxies is confirmed by larger samples,
amplification by gas coplanar with the galaxy disk would be implied.  The
100 pc-scale torus proposed by MR95 is one candidate, although it is not clear
whether such a large structure could be heated to the temperatures of 
$>$ 300 K at which water is abundant (Neufeld, Lepp \& Maloney 1995).  An
alternative source of amplification could be a molecular cloud, heated by
starlight, along the line of sight to the nucleus (Haschick \& Baan 1990).

\subsection{Mid- and Far-Infrared Properties}

The Infrared Astronomical Satellite (IRAS) mapped 98\% of the celestial
sphere at wavelengths of 12, 25, 60, and 100 $\mu m$.
Active galaxies are known to be powerful IR sources, and in many cases
the mid-IR emission of an AGN dominates its energy output and is an excellent
indicator of its bolometric luminosity.
Still, the detailed processes which determine the IR properties of
AGN are poorly understood.  The emission at mid-IR wavelengths (12 and 25
$\mu m$) are strongly related to the active nucleus, and are believed to
result from reprocessing of more energetic photons by dust near the
central engine.  The far-IR emission (60 and 100 $\mu m$), on the other hand,
can be a tracer of dust heated in star forming regions.  Ongoing
star formation is often seen in active galaxies and so may
contribute to the far-IR luminosity.  Henkel, Wouterloot \& Bally (1986) have
found that the presence of extragalactic H$_2$O masers appears to be 
correlated with the 100 $\mu m$ flux densities and the S(60)/S(100) and
S(12)/S(25) color temperatures in a sample of bright IRAS galaxies.
However, only two (NGC 1068 and NGC 3079) are megamasers, with the rest
being associated with galaxies undergoing active star formation.

Histograms of flux and power for all 4 IRAS bands are
shown in Figures 13 -- 20.  Most of the H$_2$O-detected galaxies fall
within the normal range of IR flux, though a few are unusually bright.
The statistical difference between detected and undetected galaxies in
apparent IR brightness at 12 $\mu m$ and
25 $\mu m$ is the most significant of all total galaxy properties compared,
with the detected galaxies being brighter at $>$~99\% confidence level
(Table 3).  The maser-detected galaxies are brighter at 60 and 100 $\mu m$
as well, though the significance is marginal at 100 $\mu m$.
In IR power, the distribution of the detected galaxies closely follows that of
the undetected ones in the magnitude- and distance-limited samples.
Thus, the higher apparent IR brightness of the detected galaxies may
result, at least in part, from them being nearer than the undetected
galaxies.

Because the IRAS properties of Seyfert 1s, Seyfert 2s and LINERs are
different (e.g. Miley, Neugebauer \& Soifer 1985), it is instructive to see
IR properties differentiated by AGN class.
The IRAS fluxes at 25 $\mu m$ and 60 $\mu m$ for the distance-limited
sample are shown in Figure 21 with different AGN classes plotted with
different symbols.  The LINERs are noticeably ``cooler'', on average, than
the Seyfert galaxies at these wavelengths (in the distance-limited sample,
$<$$S(25) \over S(60)$$>$ = 0.14 $\pm$ 0.02 for LINERs,
0.33 $\pm$ 0.03 for Seyfert 2s, and 0.37 $\pm$ 0.04 for 
Seyfert 1s).  The detected galaxies tend to follow the run of their
class, but three Seyfert 2s (NGC 4945, NGC 1068, and the Circinus
galaxy) and two LINERs (NGC 3079 and NGC 4258) are unusually bright.
This is consistent with the conclusion above that the detected H$_2$O masers
tend to be the apparently brightest galaxies at all IRAS bands.

A color-color diagram is shown in Figure 22 for the distance-limited
sample, with the 
ratio $S(12)/S(25)$ plotted on the abscissa and the ratio
$S(60)/S(100)$ on the ordinate.  The far-IR color is particularly
sensitive to the presence of foreground Galactic ``cirrus'' emission, which
can cause artificially small values of $S(60)/S(100)$.  The 
H$_2$O detected galaxies are all in areas relatively free of the cirrus
emission.  The entire sample
follows a pattern in which galaxies with cooler mid-IR colors have warmer
far-IR colors.  Normal galaxies follow this pattern as well (e.g. Helou 1986).
The AGN classes can be statistically separated in this figure,
with the Seyfert 2s being the coolest at short IR wavelengths, the
LINERs being the coolest at long wavelengths, and the Seyfert 1s spread
through the middle ground.
(For galaxies in the distance-limited sample, $<$$S(12) \over S(25)$$>$
is 0.61 $\pm$ 0.08, 0.39 $\pm$ 0.02 and 0.75 $\pm$ 0.05, and 
$<$$S(60) \over S(100)$$>$ is 0.53 $\pm$ 0.04, 0.66 $\pm$ 0.03 and 
0.35 $\pm$ 0.02 for Seyfert 1, Seyfert 2 and LINER galaxies, respectively.)
The maser-detected galaxies again follow their class and are unremarkable
for the most part.  However, two detected galaxies (ESO 103-G35 and
Mrk 1210) stand out as having very warm far-IR colors.  Both of these galaxies
are weak 60 $\mu m$ and 100 $\mu m$ sources with uncertainties in
log S(60)/S(100) much larger than the other sources plotted.  For ESO 103-G35
log S(60)/S(100) = 0.34 $\pm$ 0.13, and for Mrk 1210 
log S(60)/S(100) = 0.16 $\pm$ 0.08.  Notwithstanding these uncertainties, the
colors are reliably warmer than the other plotted galaxies.  Warm colors
in the far-IR are sometimes used to identify galaxies with ongoing starbursts
or an AGN, with the infrared emission originating in dust heated by hot stars
or the AGN (e.g. Rowan-Robinson and Crawford 1989).  The mid-IR colors of
ESO 103-G35 and Mrk 1210 are not unusual for the Seyfert 2 population.
A histogram showing the distribution of the S(60)/S(100) ratio is shown in
Figure 23.  The maser-detected galaxies tend to have larger S(60)/S(100)
ratios than the
undetected galaxies at a significance level of $\sim$~98\% (Table 3).
When the detected and undetected distributions of only Seyfert 2 and LINER
galaxies are considered, the significance level remains at
$\sim$~98\% (Table 3).  The difference in far-IR color is thus not influenced
by the absence of H$_2$O maser detections in Seyfert 1 galaxies.  However,
the significance level is reduced (78\%) and the effect marginal when only the
Seyfert 2s are included.

In Figure 24, the ratio S(25)/S(60) is plotted against the power in the
25 $\mu m$ band for the distance-limited
sample.  The general trend for both H$_2$O-detected and undetected galaxies
is for the color to become warmer as the IR power
increases.  The LINERs again tend to be intrinsically weaker and
cooler than the Seyferts.  The color of the maser-detected galaxies shows
considerable dispersion, with two being particularly warm (Mrk 1210 and
ESO 103-G35 -- the same galaxies as were unusually warm in the S(60)/S(100)
ratio) and two being unusually cool (NGC 3079 and NGC 4945).
The megamaser galaxy NGC 1052 is the warmest of all observed
LINERs.  Nevertheless, the detected galaxies, as a group, are not strongly
associated with warm S(25)/S(60) IRAS ratios (Table 3).  A histogram
showing the distribution of S(25)/S(60) for the detected and undetected
galaxies in the distance-limited sample is shown in Figure 25.
In summary, the galaxies detected as H$_2$O masers tend to be the apparently
brightest at mid- and far-infrared wavelengths and to have unusually warm
far-infrared colors.

\section{Nuclear properties}

\subsection{AGN Class}

The most striking trend from our survey is that no Seyfert 1s were detected.
Ten of the masers are found in Seyfert 2 galaxies and six are in LINERs
(including the galaxy TXFS 2226-184 (Koekemoer et al. 1995),
which is not a member of the samples discussed in this paper).
The fraction of H$_2$O-detected LINERs in the distance-limited sample --
0.075 ($5 \over 67$) -- and magnitude-limited sample -- 0.079 ($5 \over 63$) --
is similar to that of Seyfert 2s -- 0.071 ($10 \over 141$) and 0.071
($8 \over 112$), respectively.  However,
the LINERs in our sample tend to be closer than the Seyfert 2s. The fraction
of detected Seyfert 2s in the sensitivity-limited sample -- 0.14
($10 \over 73$) -- is thus somewhat larger than, but comparable to, that of
LINERs -- 0.096 ($5 \over 52$).

The detected fraction of Seyfert 2s and LINERs together are
0.072 $\pm$ 0.019 ($15 \over 208$), 0.074 $\pm$ 0.021 ($13 \over 175$) and
0.069 $\pm$ 0.018 ($15 \over 215$) in the distance-limited,
magnitude-limited, and combined distance- and magnitude-limited samples,
respectively.  If Seyfert 1s have the same detection probabilities, the
random chances of detecting 0 out of the 57 (distance-limited), 54
(magnitude-limited), and 70 (combined distance- and magnitude-limited) samples
are only 0.014 $^{+ 0.030}_{- 0.010}$, 0.015 $^{+ 0.035}_{- 0.011}$, and
0.006 $^{+ 0.018}_{- 0.005}$, respectively.
This difference in detection rates cannot be ascribed to a larger
mean distance to the Seyfert 1s, which are located at the same mean
distance (53.0 $\pm$ 3.2 Mpc) as the Seyfert 2s (52.9 $\pm$ 1.9 Mpc) for the
distance-limited sample, though the LINERs in the distance-limited sample 
are nearer (37.3 $\pm$ 3.0 Mpc), as mentioned earlier.
These statistics show that the incidence of detectable H$_2$O
megamasers in Seyfert 1 galaxies is much smaller than in Seyfert 2s and 
LINERs.

The absence of detections in Seyfert 1 galaxies indicates either
that these galaxies
do not have molecular gas with appropriate conditions to mase, or that the
masers in these galaxies are beamed away from our line of sight. 
The first possibility would violate the strongest version of the unified
scheme, in which the populations of Seyfert 1s and 2s are identical except for angle of view.  It might, however, be consistent with scenarios in which
Seyfert 2s evolve into 1s or vice versa (e.g. Miller 1995).  Turning to the
second possibility,
Seyfert 2 galaxies are thought to contain Seyfert 1 nuclei obscured by a
pc-scale thick disk or torus (e.g. Antonucci 1993).
A reasonable scenario is that the masers originate in gas associated with
the torus and are beamed in its plane.  VLBA observations
of NGC 4258 (Miyoshi et al. 1995) show that the
masers are found in a thin edge-on disk, so if other H$_2$O megamasers
have a similar structure, our detection of only type 2 Seyferts indicates that
the thick torus and thin disk tend to be coplanar.
One possibility is that the thick torus and thin
disk occupy the same radial range from the nucleus, with the gas density
in the thick torus increasing towards the central plane.
Another possibility is that the thin disk is nearer to the nucleus than the
torus. The thin disk might then flare out with increasing distance
from the nucleus and gradually transform into a thick disk.
Recent high resolution observations of NGC 1068 (Greenhill et al. 1996)
suggest that some maser emission is associated with the inner edge of the torus.

\subsection{Nuclear V Magnitude}

It is difficult to make statistical conclusions based on optical photometric
properties of AGN because there is no data set available which
measures only the nonstellar nuclear component
of the V band emission.  Both Huchra and V\'eron-Cetty \& V\'eron
list V magnitudes which are taken with small aperture measurements (i.e.
$\lesssim$~15\arcsec) when available.  For a given aperture size and nuclear
luminosity, the fractional unwanted contribution from stars in
the host galaxy is a function of the galaxy's distance.
The accumulation of data from a number of different surveys
with varying apertures also complicates the picture.  Nonetheless, it is
useful to check the distribution of nuclear magnitudes using the available data.
The magnitudes were taken from the Huchra catalog when given there,
otherwise the values were taken from the V\'eron-Cetty \& V\'eron catalog.

Figures 26 and 27 show the distributions of apparent and absolute
``nuclear'' V magnitude.  There is no significant difference between the 
maser-detected and undetected galaxies.

\subsection{Radio Flux and Luminosity}

The radio emission from Seyfert and LINER galaxies is, in general, weak
and often dominated by the circumnuclear (100 pc -- kpc scale) regions.
To ensure that only
small-scale ($\lesssim$ kpc) flux is considered, VLA data at 6 cm
with the A and B configurations are considered exclusively
(e.g. Unger et al., 1987; Ulvestad \& Wilson, 1984a, 1984b, 1989).
Such radio measurements are available for 25\% of the distance-limited
sample.  Radio fluxes at 1.3 cm (the wavelength of the 6$_{16}$ -- 5$_{23}$
H$_2$O transition) are available for even fewer
sources, and are not directly addressed here.  Although about
10\% of Seyferts have a flat radio spectrum nuclear source at centimeter
wavelengths (probably due to synchrotron
self-absorption), the majority are steep spectrum sources with spectral
index $\alpha$ $\simeq$~0.7 between 6 cm and 20 cm, where S $\propto$
$\nu^{-\alpha}$ (e.g. Ulvestad
\& Wilson, 1984b).  We shall use the 6 cm flux and luminosity as indicators
of nuclear radio continuum output.

Figures 28 and 29 show histograms of the total nuclear radio flux and power
at 6 cm.  The radio fluxes of the H$_2$O-detected galaxies are stronger than
the undetected ones at the $\sim$~95 -- 98\% confidence level (Table 3).
The radio power of the detected sources may be stronger as well, but with
weaker ($<$92\%) statistical significance (Table 3).  The statistics are
limited by the small number of detected masers with nuclear radio continuum
fluxes.

The observed H$_2$O maser emission might result from amplification of a 1.3 cm
nuclear radio source.  Such amplification can only occur if the ``seed''
source of radio continuum radiation is behind and smaller than the masing
clouds.
If the masers arise in a parsec-scale region or smaller (e.g. Miyoshi et al.
1995) then any radio flux spatially resolved by the VLA cannot contribute
to the seed emission; only an unknown fraction of the unresolved radio flux
is relevant.  For example, the best VLA resolution at 6 cm is
$\sim$~0.4\arcsec, which corresponds to a linear scale of $\sim$~20 pc (much
larger than the expected scale of the maser emission) for a galaxy
at a distance of 10 Mpc.

The source of the compact radio luminosity in Seyfert galaxies is not fully
understood, but
an interesting clue is that when the compact structure is resolved, it is
seen as a linear or jet structure in most cases (Ulvestad \& Wilson, 1984b).
Such a jet is believed to
be collimated by the accretion disk, so the possible trend for masers to be
associated with more powerful radio continuum sources may indicate that
the masers are related to currently active accretion and jet formation.

\subsection{[OIII] Flux and Luminosity}

Nuclear [OIII] fluxes for the observed galaxies were compiled primarily from
Dahari \& De Robertis (1988), Ho, Filippenko \& Sargent (1993), Oliva et al.
(1994),
Stauffer (1982), and Whittle (1992).  No corrections have been made for
extinction, as this is often unknown.

The distributions of [OIII]$\lambda$5007 fluxes and luminosities are shown in
Figures 30 and 31.  Because [OIII] is excited
primarily by the nuclear ultraviolet continuum, it can be used as a measure of
the power of the AGN at those wavelengths.  Large-aperture
[OIII] measurements are used when available; otherwise smaller aperture
measurements are listed.  Unfortunately, [OIII]$\lambda$5007 fluxes are
available for only eight maser-detected galaxies.  Nevertheless, the
available data show that the H$_2$O-detected galaxies are well distributed
among the range
of [OIII] luminosity (and flux), suggesting the UV output of the AGN is
not an important parameter for H$_2$O maser production (cf. Table 3).

A plot of the [OIII]/H$\beta$ ratio against the [OIII] luminosity for
each class of AGN in the distance-limited sample is shown in Figure 32.
The three classes occupy distinct regions
of this plot, with the LINERs being systematically weakest in [OIII]
luminosity, and Seyfert 2s having larger [OIII]/H$\beta$ ratios than Seyfert
1s (the H$\beta$ flux of which includes both narrow and broad components).
The detected galaxies occupy typical locations on this
diagram.  The two H$_2$O-detected LINERs plotted in Figure 32 seem to
have higher [OIII] 
luminosities than the undetected LINER galaxies, but two other LINERs
have low values (the log of the [OIII]$\lambda$5007 luminosity is
40.5 for NGC 4258, and 40.0 for
NGC 3079) but are not plotted because they lack a reliable nuclear H$\beta$
measurement.

\subsection{X-ray Properties}

Details of the pumping mechanism for H$_2$O megamasers are not yet understood,
but the association of megamasers with AGN suggests that the nonstellar
nuclear emission
is the ultimate energy source.  Neufeld, Maloney \& Conger (1994)
present a model in which the X-ray source in the AGN heats the nuclear gas,
giving rise to a molecular layer in which powerful, collisionally-pumped H$_2$O
masers can exist.  In their model, the true luminosity of the maser increases
in proportion to the area of gas illuminated by the X-ray source.  
Neufeld \& Maloney (1995) show that this X-ray heating model fits the
observations of the masing disk in NGC 4258 (Miyoshi et al. 1995), and
they suggest the possibility of even more powerful masers based on reasonable
X-ray parameters.

Observations of hard (2 -- 10 keV) X-rays have so far been reported in
nine of the maser-detected galaxies.  In NGC 1068 the observed 2 -- 10 keV 
X-ray luminosity is only 1.4$\pm$0.5 $\times$
10$^{41}$ erg s$^{-1}$ (Koyama et al. 1989, corrected to H$_0$ = 75 km
s$^{-1}$ Mpc$^{-1}$), which may be understood as the
reflected component of a nuclear X-ray source with an intrinsic
luminosity of 10$^{43}$ -- 10$^{44}$ erg s$^{-1}$ (Koyama et al. 1989) and
which is completely obscured by a column of $\gtrsim$~10$^{25}$ cm $^{-2}$
(e.g. Mulchaey, Mushotzky \& Weaver 1992).
The Circinus galaxy also exhibits a reflection-dominated X-ray spectrum (Matt
et al. 1996).
Matt et al. (1996) estimate the X-ray luminosity of the hidden nucleus to
be $\sim$ 4 $\times$ 10$^{41}$/($\Omega/2\pi$) erg s$^{-1}$ ($\Omega$
being the solid angle subtended at the nucleus by the reflecting matter),
and the absorbing column at $>$ 10$^{24}$ cm $^{-2}$.
The estimated 2 -- 10 keV luminosities (corrected for absorption) and 
the equivalent hydrogen column densities inferred
from the X-ray spectra are listed in Table 4,
and plotted against the observed maser luminosities in Figures 33 and 34,
respectively.  No trends are apparent in either of these diagrams.

Awaki (1996) has summarized ASCA results from a sample of 17
Seyfert 2 galaxies selected according to [OIII] brightness, polarized
broad line detection and previous X-ray observations.  Figure 2
in Awaki (1996) shows the distribution of absorbing column for
this sample plus 5 additional ASCA-observed Seyfert 2 galaxies, and we have
reproduced this diagram as Figure 35 here.  Twenty one out of the 22
ASCA-observed Seyfert 2 galaxies were included in our H$_2$O maser survey
(the unobserved galaxy is NGC 6552).  Five out of these 21 
are H$_2$O-detected galaxies, and they are shaded in
Figure 35.  It is interesting to note that, among galaxies in this sample,
3 of the 5 which have been searched for H$_2$O maser emission and have 
log N$_H$ $>$ 10$^{24}$
cm$^{-2}$ are maser-detected sources, while only 2 of the 16 with 
log N$_H$ $<$ 10$^{24}$ have been detected as H$_2$O sources.  A 2 $\times$ 2
contingency table with these results gives a $\chi$$^2$ of 4.7, indicating
that, among this sample, the masers are found in sources with larger column
densities at a significance level of 97\%.
This result suggests that H$_2$O megamaser emission in Seyfert 2s is
preferentially associated with galaxies having high X-ray absorbing columns.
The non-detection of Seyfert 1 galaxies in H$_2$O maser emission may then
be understood in terms of their lack of a large column of cool gas along the 
line of sight to the nucleus.  We caution, however, that this possible
association between H$_2$O megamaser emission and high X-ray column may be
subject to selection effects.  For example, the three galaxies with
N$_H$ $>$ 10$^{24}$ cm$^{-2}$ and detected in H$_2$O emission (NGC 1068, 
NGC 4945 and the Circinus galaxy) are amongst the nearest of the sample
of galaxies searched for H$_2$O emission; it may be that detection of
the reflected component of hard X-rays is only possible in the
nearest Seyfert 2s.  It is hoped that ASCA
observations will be available for the remaining seven
H$_2$O-detected sources in the near future, and will allow further
investigation of the relationships between H$_2$O maser and hard X-ray
emission.

\section{Results and Conclusions}

The strongest effect that we have identified is that H$_2$O emission is only
found in Seyfert 2s and LINERs.  No H$_2$O emission is detected from
Seyfert 1s.  The detection rates of Seyfert 2s and LINERs are similar; this
finding constitutes a strong argument that at least some nuclear LINERs are
AGNs rather than starbursts.
The lack of detections of Seyfert 1s indicates that either they do not have
molecular gas in their nuclei with physical conditions appropriate to
produce H$_2$O masers, or the masers are beamed away from Earth.  Our result
is completely consistent with the notion that the masers are beamed in the
plane of the thick nuclear torus postulated to hide the Seyfert 1 nucleus
in Seyfert 2 galaxies.

With good confidence ($>$ 95\%), we also preferentially detect H$_2$O
emission from the nearer galaxies and from those which are apparently brighter
at mid-infrared (12 and 25 $\mu m$), far-infrared (60 $\mu m$), and
short centimeter radio (6 cm) wavelengths, and have warm far-infrared
IRAS colors.  At somewhat lower confidence
level ($\sim$ 90\%), we find H$_2$O emission preferentially from galaxies
more luminous in 6 cm radio emission.

The general tendency to detect nearer and apparently brighter galaxies is in
accordance with general astronomical expectations (cf. Section 2).  
The strong tendency to detect the apparently brightest galaxies at
mid-infrared (12 $\mu m$ and 25 $\mu m$) wavelengths, where most of the nuclear
bolometric luminosity of Seyfert 2s may be emitted, suggests a correlation 
between H$_2$O megamaser and nuclear bolometric luminosities.  Further,
the tendency to detect galaxies with warm far-infrared IRAS colors
suggests that galaxies with nuclear far-infrared emission dominant over disk 
emission tend to be more easily detectable.

The most significant correlation of H$_2$O maser detectability with an 
intrinsic luminosity is that with nuclear centimeter wavelength radio power.
As the radio continuum emission is generally extended on the hundreds pc 
scale, it is
unlikely that this correlation simply reflects one between maser luminosity
and ``seed'' photon availability.  Radio jets and lobes are one of the best
indicators of current nuclear activity in Seyfert 2s and the relationship 
with maser emission may just reflect a mutual dependence on the level of
current activity.  Alternatively, a distinct physical process may be involved.
For example, radio ejecta are undoubtedly associated with shocks and it is
possible that the shock-heated gas may favor H$_2$O maser emission.

There is a trend for H$_2$O megamaser emission in Seyfert 2s to be 
preferentially associated with galaxies exhibiting a very high X-ray absorbing
column (N$_H$ $>$ 10$^{24}$ cm$^{-2}$).  Our data are consistent with the
notion that the probability of detecting H$_2$O maser emission from Seyfert
galaxies increases with increasing column of cool gas to the nucleus, from
Seyfert 1s through NLXGs to Seyfert 2s.

\acknowledgements
We have made extensive use of the NASA/IPAC Extragalactic
Database (NED) which is operated by the Jet Propulsion Laboratory,
California Institute of Technology, under contract with the National
Aeronautics and Space Administration.
We thank Eric Feigelson for assistance with statistical analysis and for 
providing the ASURV Rev. 1.2 software (LaValley, Isobe \& Feigelson 1992)
which implements the methods presented in Feigelson \& Nelson (1985).
We also thank Kim A. Weaver for valuable discussions on X-ray properties.
This research was supported by NASA through grants NAGW-3268, 
HST GO grant 3724 and by NSF through grant AST-9527289.  
C.H. acknowledges support from NATO grant SA.5-2-05 (CRG. 960086) 318/96.

\newpage

\newpage
\pagestyle{empty}
\begin{table}
\begin{tabular}{lcccccccc}
\multicolumn{9}{c}{Table 2} \\
\hline
\multicolumn{9}{c}{Properties of the Statistical Samples} \\
\multicolumn{1}{c}{Sample} &
\multicolumn{1}{c}{Total} &
\multicolumn{1}{c}{Seyfert 1} &
\multicolumn{1}{c}{Seyfert 2} &
\multicolumn{1}{c}{LINER} &
\multicolumn{1}{c}{Unspecified} &
\multicolumn{1}{c}{Huchra} &
\multicolumn{1}{c}{V\'eron} &
\multicolumn{1}{c}{Both} \\
\multicolumn{1}{c}{(1)} &
\multicolumn{1}{c}{(2)} &
\multicolumn{1}{c}{(3)} &
\multicolumn{1}{c}{(4)} &
\multicolumn{1}{c}{(5)} &
\multicolumn{1}{c}{(6)} &
\multicolumn{1}{c}{(7)} &
\multicolumn{1}{c}{(8)} &
\multicolumn{1}{c}{(9)} \\
\hline
Detected Masers      &  15  &  0  &  10  &  5  &  0  &  13  &  12  &  11  \\
Magnitude-limited    & 241  & 54  & 112  & 63  & 12  & 186  & 206  & 152  \\
Distance-limited     & 278  & 57  & 141  & 67  & 13  & 207  & 238  & 168  \\
Sensitivity-limited  & 165  & 30  &  73  & 52  & 10  & 127  & 147  & 110  \\
All in above samples & 299  & 70  & 149  & 67  & 13  & 226  & 255  & 183  \\
\tablecomments{All 15 maser-detected galaxies are included in the
distance- and sensitivity-limited samples, but only 13 are included in the
magnitude-limited sample, Mrk 1 and ESO 103-G35 being too faint.
The columns show (1) the sample, (2) the total number of
galaxies for that sample (note that one maser-detected galaxy -- 
TXFS 2226-184 --
is not in any of our samples), (3) the number of Seyfert 1 galaxies, (4) the
number of Seyfert 2 galaxies, (5) the number of LINERs, (6) the number of
AGNs included which have no detailed classification available, (7) the
number from the Huchra catalog, (8) the number from the V\'eron \&
V\'eron-Cetty catalog, and
(9) the number common to both the Huchra and V\'eron \& V\'eron-Cetty
catalogs.}
\end{tabular}
\end{table}

\newpage

\begin{table}
\begin{tabular}{lrrc}
\multicolumn{4}{c}{Table 4} \\
\hline
\multicolumn{4}{c}{Hard X-ray Observations of H$_2$O Megamaser Galaxies} \\
\multicolumn{1}{c}{Source} &
\multicolumn{1}{c}{N$_H$} &
\multicolumn{1}{c}{L$_X$\tablenotemark{a}} &
\multicolumn{1}{c}{Ref\tablenotemark{b}} \\
\multicolumn{1}{c}{} &
\multicolumn{1}{c}{(cm$^{-2}$)} &
\multicolumn{1}{c}{(erg s$^{-1}$)} \\
\hline
NGC 1068    & $>$ 10$^{25}$& 10$^{43}$ - 10$^{44}$& 1 \\
NGC 1386    & 5 $\pm$ 3 $\times$ 10$^{23}$ & - & 2 \\
Mrk 1210    & 1.1 $\pm$ 0.8 $\times$ 10$^{23}$ & - & 2 \\
NGC 3079    & 1.6 $^{+2.4}_{-1.6}$ $\times$ 10$^{22}$& 5.8 $\times$ 10$^{40}$ &
 3 \\
NGC 4258    & 1.5 $\pm$ 0.2 $\times$ 10$^{23}$&4 $\pm$ 1$\times$ 10$^{40 }$&4 \\
NGC 4945    & 5.4 $\pm$ 0.5 $\times$ 10$^{24}$& 1 $\times$ 10$^{42}$ & 5 \\
Circinus    & $>$ 10$^{24}$&4 $\times$ 10$^{41}$ - 4 $\times$ 10$^{43}$& 6 \\
NGC 5506    & 2.8 $\pm$ 0.5 $\times$ 10$^{22}$& 1.7 $\times$ 10$^{43}$ & 7 \\
ESO 103-G35 & 1.35 $^{+.33}_{-.23}$ $\times$ 10$^{23}$&
3.3 $\times$ 10$^{43}$ & 7 \\

\tablenotetext{a}{2 -- 10 keV luminosity, corrected for photoelectric
absorption.}
\tablenotetext{b}{(1) Koyama et al. (1989); (2) Awaki (1996);
(3) Serlemitsos et al. (1996); (4) Makishima et al. (1994);
(5) Iwasawa et al. (1993); (6) Matt et al. (1996); (7) Turner \& Pounds (1989)}
\end{tabular}
\end{table}

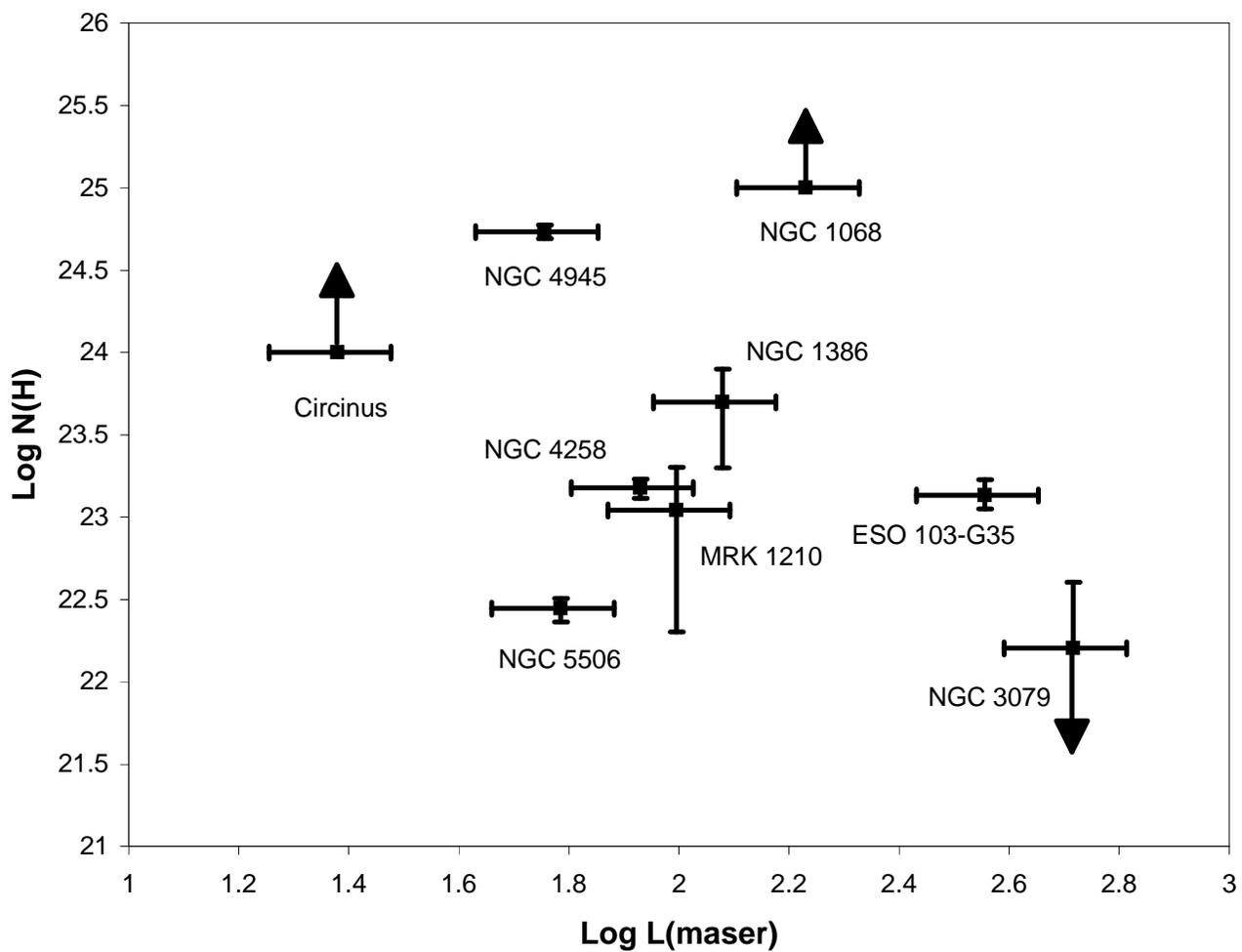
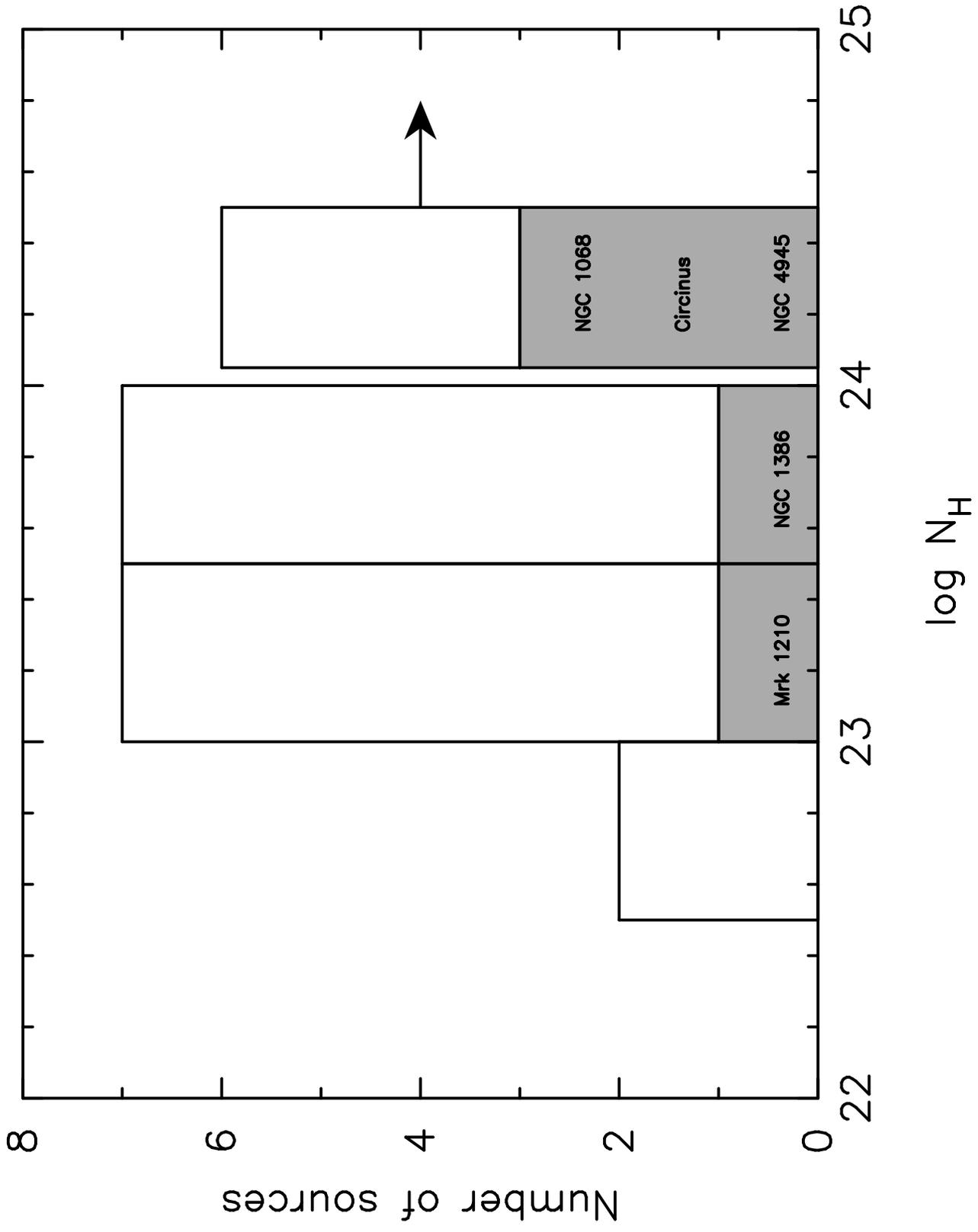
\end{document}

\pagestyle{empty}
\begin{planotable}{lrrrrrrrrrrrrrrr}
\tablenum{1-a}
\tablecaption{Properties of Observed Galaxies}
\tablehead{
\colhead{Source} &
\colhead{M} &
\colhead{D} &
\colhead{S} &
\colhead{V} &
\colhead{Dist} &
\colhead{Class} &
\colhead{T} &
\colhead{Inc} &
\colhead{$m_B$} &
\colhead{$M_B$} &
\colhead{$m_V$} &
\colhead{$M_V$} &
\colhead{S$_{5007}$} &
\colhead{S$_{H\beta}$} &
\colhead{log P$_{5007}$} \\
\colhead{} &
\colhead{} &
\colhead{} &
\colhead{} &
\colhead{(km s$^{-1}$)} &
\colhead{(Mpc)} &
\colhead{} &
\colhead{} &
\colhead{(degrees)} &
\colhead{} &
\colhead{} &
\colhead{} &
\colhead{} &
\colhead{(cgs)} &
\colhead{(cgs)} &
\colhead{(cgs)} \\
\colhead{(1)} &
\colhead{(2)} &
\colhead{(3)} &
\colhead{(4)} &
\colhead{(5)} &
\colhead{(6)} &
\colhead{(7)} &
\colhead{(8)} &
\colhead{(9)} &
\colhead{(10)} &
\colhead{(11)} &
\colhead{(12)} &
\colhead{(13)} &
\colhead{(14)} &
\colhead{(15)} &
\colhead{(16)} \\
}

\startdata

Mrk 334       &        1&         1&          1&     6600&       84.4&        2&      99&     45&   14.38&         &    14.62&   -19.99&      4.9&      3.3&    42.61    \nl
Mrk 335       &        1&         0&          0&     7688&       99.2&        1&       0&     27&   13.75&         &    13.85&   -21.11&       23&       95&    43.42    \nl
Mrk 937       &        0&         0&          0&     8851&      115&        2&        &      0&      16&         &     13.8&   -21.47&         &         &             \nl
Mrk 938       &        1&         1&          0&     5931&       75.0&        2&      99&     72&      14&         &    13.52&   -20.84&      1.1&     0.15&    41.86    \nl
0017-5133     &        1&         1&          0&     6586&       86.2&       L &      -1&     74&   14.48&   -20.32&         &         &         &         &             \nl
Mrk 945       &        0&         0&          0&     4431&       54.9&       O &        &      0&      15&         &    14.54&   -19.14&         &         &             \nl
NGC 224       &        0&         0&          0&     -300&        0.7&       O &       3&     76&    4.36&   -20.89&    10.57&   -13.66&         &         &             \nl
NGC 235A      &        1&         1&          0&     6664&       87.7&        2&      -5&     59&   14.08&   -20.91&     14.2&   -20.49&         &         &             \nl
Mrk 348       &        1&         1&          1&     4467&       56.5&        2&       0&     16&    13.9&   -19.81&    14.59&   -19.16&       42&        4&    43.20    \nl
Mrk 352       &        0&         1&          1&     4456&       55.9&        1&      -2&     50&    14.8&   -19.23&    14.81&   -18.91&      4.9&       22&    42.26    \nl
ESO 079-G16   &        1&         1&          0&     6036&       78.8&        2&       0&     72&   14.31&   -20.84&    14.38&   -20.08&         &         &             \nl
NGC 404       &        1&         1&          1&      -48&        2.4&       L &      -3&      0&   11.21&   -15.98&    11.72&   -15.18&      2.5&     1.79&    39.24    \nl
Tol 0109-383  &        1&         1&          0&     3496&       43.2&        2&       0&     75&   13.73&   -19.77&    14.12&   -19.05&       42&      9.1&    42.97    \nl
IC 1657       &        1&         1&          0&     3552&       44.4&        2&       4&     85&   13.15&   -21.03&         &         &         &         &             \nl
NGC 454       &        1&         1&          0&     3645&       46.9&        2&      99&       &   13.12&         &    13.58&   -19.76&         &         &             \nl
{\bf Mrk 1}   &        0&         1&          1&     4839&       61.0&        2&       0&     45&   15.01&   -19.46&    14.96&   -18.95&       60&      5.4&    43.42    \nl
Mrk 565       &        1&         1&          0&     5446&       69.0&       U &      -2&     38&   13.35&   -20.97&     13.5&   -20.67&         &         &             \nl
0120-0214     &        0&         1&          0&     4730&       59.4&        2&       0&     71&   14.88&   -19.87&    14.95&   -18.90&         &         &             \nl
NGC 513       &        1&         1&          0&     5859&       76.4&        2&       0&     67&    13.9&   -21.22&     13.4&   -20.99&         &         &             \nl
NGC 526A      &        0&         1&          0&     5762&       74.5&        2&      -2&     60&   14.76&   -19.74&     14.6&   -19.74&       27&      2.1&    43.25    \nl
Mrk 993       &        1&         1&          1&     4658&       58.9&        2&       1&     75&   14.39&   -20.29&    13.96&   -19.87&        3&     0.52&    42.09    \nl
Mrk 359       &        1&         1&          1&     5007&       63.8&        2&      -2&     43&   14.16&         &    14.22&   -19.79&       11&      3.5&    42.72    \nl
0129-3323     &        1&         1&          0&     4958&       63.6&        2&       4&     31&   13.67&   -20.46&         &         &         &         &             \nl
NGC 600       &        1&         1&          1&     1842&       22.9&       L &       6&     33&   12.92&   -19.09&    12.57&   -19.23&         &         &             \nl
NGC 591       &        1&         1&          1&     4547&       57.6&        2&       0&     34&   13.89&   -20.16&     13.9&   -19.89&       23&      2.3&    42.95    \nl
NGC 613       &        1&         1&          0&     1475&       16.6&       U &       4&       &   10.74&   -20.56&    12.71&   -18.38&         &         &             \nl
Mrk 1158      &        0&         0&          0&     4585&       58.0&       O &        &     32&   15.03&   -19.01&    14.96&   -18.84&      8.8&      3.7&    42.54    \nl
Mrk 573       &        1&         1&          0&     5096&       64.7&        2&      -1&     20&   13.68&   -20.47&    14.07&   -19.97&      160&       13&    43.90    \nl
0147-0740     &        0&         1&          0&     5306&       67.7&        2&      -4&       &    15.5&         &    15.62&   -18.51&         &         &             \nl
0152+0622     &        1&         1&          1&     5164&       65.7&        2&       3&     41&   14.18&   -20.21&     14.5&   -19.57&         &         &             \nl
IC 184        &        0&         1&          0&     5382&       68.8&        2&       1&     62&   14.66&         &     14.1&   -20.07&         &         &             \nl
NGC 788       &        1&         1&          1&     4078&       51.6&        2&       0&     35&      13&   -20.75&    12.76&   -20.79&       15&      1.5&    42.67    \nl
ESO 153-IG16  &        0&         1&          0&     5798&       77.9&        2&       6&     65&   15.04&   -19.88&    15.64&   -18.80&         &         &             \nl
ESO 153-G020  &        1&         1&          0&     5917&       78.2&        2&       3&       &   14.03&   -20.95&    14.31&   -20.14&         &         &             \nl
NGC 824       &        1&         1&          0&     5836&       76.0&       U &       4&       &   14.11&   -20.45&     14.1&   -20.28&         &         &             \nl
4C 35.03      &        0&         0&          0&    11173&      149&       R &      -5&       &    14.1&   -22.06&         &         &         &         &             \nl
NGC 863       &        1&         0&          0&     7910&      103&        1&       1&     25&   13.85&   -21.38&    13.81&   -21.24&      5.3&        6&    42.82    \nl
Ark 79        &        1&         1&          1&     5157&       66.4&        1&       1&     82&    14.1&   -21.07&    13.84&   -20.25&      3.2&      0.4&    42.22    \nl
NGC 931       &        1&         1&          1&     4927&       64.1&        1&       4&     81&   14.46&   -21.79&    14.74&   -19.28&       13&       14&    42.80    \nl
Mrk 1044      &        1&         1&          0&     4932&       63.2&        1&        &     32&    14.5&         &    14.29&   -19.70&      5.4&       33&    42.40    \nl
IC 1816       &        1&         1&          0&     5037&       67.3&        2&       2&     32&   13.83&   -20.46&     13.6&   -20.52&         &         &             \nl
NGC 1019      &        1&         0&          0&     7251&       94.7&        1&       4&     25&   14.34&   -20.69&    14.95&   -19.91&         &         &             \nl
{\bf NGC 1052}&        1&         1&          1&     1470&       17.8&       L &      -5&     47&   11.41&   -19.90&    12.31&   -18.94&    11.74&     5.93&    41.65    \nl
{\bf NGC 1068}&        1&         1&          1&     1136&       14.4&        2&       3&     29&    9.61&   -21.32&    10.83&   -19.96&     1995&      169&    43.69    \nl
NGC 1097      &        1&         1&          1&     1275&       14.5&        1&       3&     49&   10.23&   -20.89&    12.78&   -18.03&         &         &             \nl
Mrk 372       &        0&         0&          0&     9300&      123&        2&       0&     38&    14.9&   -21.07&    14.81&   -20.61&       13&       10&    43.36    \nl
Mrk 1058      &        0&         1&          1&     5138&       66.4&        2&       1&     65&   15.38&   -19.52&    14.54&   -19.55&      5.6&      0.4&    42.46    \nl
NGC 1125      &        1&         1&          0&     3297&       41.6&        2&       0&     62&   13.43&         &    13.13&   -19.96&         &         &             \nl
NGC 1144      &        1&         0&          0&     8647&      114&        2&        &     52&   13.78&   -22.05&    14.41&   -20.85&      4.3&     0.33&    42.81    \nl
Mrk 1066      &        1&         1&          1&     3605&       45.9&        2&      -1&     42&   13.64&   -20.26&    13.96&   -19.34&       24&      5.5&    42.78    \nl
NGC 1167      &        1&         1&          1&     4945&       64.1&       L &      -3&     36&   13.38&   -21.26&    12.77&   -21.25&     6.95&     2.51&    42.53    \nl
H0258-015     &        1&         1&          1&     4017&       51.2&        1&        &       &   14.36&         &    14.36&   -19.17&         &         &             \nl
NGC 1194      &        1&         1&          0&     3957&       50.5&        1&      -1&     58&   13.83&         &    14.18&   -19.32&         &         &             \nl
NGC 1241      &        1&         1&          0&     4030&       51.6&        2&       3&     54&   12.84&   -21.47&     12.7&   -20.85&        1&  $<$.03 &    41.50    \nl
Mrk 1073      &        1&         1&          0&     6991&       92.1&        2&       3&     32&    13.7&   -21.90&    14.18&   -20.62&       22&      2.8&    43.34    \nl
0314-2303     &        1&         1&          0&     3880&       50.7&       L &       2&     35&   13.17&   -20.46&         &         &         &         &             \nl
NGC 1275      &        1&         1&          0&     5260&       68.6&        1&      99&     40&   12.64&         &    12.55&   -21.61&      283&     74.7&    44.20    \nl
NGC 1301      &        1&         1&          0&     3944&       51.7&       L &       3&     80&    14.1&   -20.59&         &         &         &         &             \nl
NGC 1326      &        1&         1&          1&     1362&       16.9&       L &      -1&     43&   11.41&   -19.75&         &         &         &         &             \nl
Mrk 607       &        1&         1&          1&     2716&       37.7&        2&       1&     76&   13.32&         &       14&   -18.88&       14&      1.4&    42.38    \nl
\tablebreak
MCG -2-9-40   &        0&         1&          0&     4495&       58.3&        2&        &     74&      15&         &    14.15&   -19.66&         &         &             \nl
Mrk 612       &        0&         1&          0&     6041&       79.2&        2&       0&     59&    15.5&         &    14.25&   -20.22&       18&      1.8&    43.12    \nl
NGC 1358      &        1&         1&          1&     4013&       51.9&        2&       0&     42&   13.04&   -20.86&    13.05&   -20.51&       13&      1.7&    42.62    \nl
NGC 1365      &        1&         1&          1&     1662&       16.9&        2&       3&     58&   10.32&   -21.21&    12.95&   -18.19&         &         &             \nl
{\bf NGC 1386}&        1&         1&          1&      924&       16.9&        2&      -1&     67&   12.09&   -19.02&    12.84&   -18.30&       80&      5.3&    42.44    \nl
III Zw 55     &        0&         0&          0&     7527&       99.6&        2&      -5&      0&    15.4&         &    15.22&   -19.75&      5.9&        1&    42.83    \nl
0339-2124     &        1&         1&          0&     4341&       56.6&        1&       1&     46&   12.83&   -21.15&    13.68&   -20.07&         &         &             \nl
NGC 1433      &        0&         0&          0&     1075&       11.6&       O &       2&       &   10.75&   -19.68&         &         &         &         &             \nl
NGC 1566      &        1&         1&          1&     1496&       13.4&        1&       4&     38&   10.13&   -20.48&    13.17&   -17.47&       30&       18&    41.81    \nl
0425-0440     &        0&         1&          0&     4647&       61.5&        2&        &       &    14.6&         &         &         &         &         &             \nl
ESO 202-G23   &        1&         1&          0&     4947&       66.2&        2&      99&     32&   13.48&         &     14.9&   -19.19&         &         &             \nl
NGC 1598      &        0&         0&          0&     5106&       68.4&       O &       5&       &   13.88&   -20.72&         &         &         &         &             \nl
0438-0828     &        1&         1&          0&     4527&       60.1&        2&       0&       &      14&         &         &         &     0.77&         &    41.52    \nl
NGC 1672      &        0&         0&          0&     1350&       14.5&       O &       3&     34&   10.28&   -20.56&    12.23&   -18.58&         &         &             \nl
NGC 1667      &        1&         1&          1&     4547&       60.6&        2&       5&     38&   12.77&   -21.49&    12.86&   -21.04&     6.04&     0.49&    42.42    \nl
H0448-041     &        1&         1&          1&     3897&       51.8&        1&        &       &   14.18&         &    14.18&   -19.38&         &         &             \nl
NGC 1685      &        1&         1&          1&     4527&       60.3&        2&       0&     47&    14.5&         &    15.18&   -18.71&         &         &             \nl
0450-0317     &        1&         1&          0&     4737&       63.2&        2&        &       &    13.6&         &       15&   -18.99&        4&     0.26&    42.27    \nl
0456+04       &        1&         1&          1&     4723&       63.0&        1&       1&     57&   13.83&   -20.90&    14.41&   -19.57&      4.8&      8.5&    42.35    \nl
ESO 033-G02   &        0&         1&          0&     5426&       73.5&        2&        &       &        &         &     14.6&   -19.71&         &         &             \nl
0459+0328     &        0&         1&          1&     4497&       60.1&        1&        &       &   14.75&         &       15&   -18.88&         &         &             \nl
NGC 1808      &        1&         1&          1&      989&       10.8&        2&       1&     55&   10.76&   -19.72&    12.55&   -17.62&         &         &             \nl
UGC 3255      &        0&         1&          1&     5689&       76.3&        2&       3&     81&   15.35&   -20.48&     15.6&   -18.79&         &         &             \nl
ESO 362-G8    &        1&         1&          0&     4785&       64.5&        2&       0&     62&    13.6&   -20.52&    12.65&   -21.38&         &         &             \nl
H0510+031     &        0&         1&          1&     4917&       66.0&        1&        &       &    14.8&         &     14.8&   -19.28&         &         &             \nl
Ark 120       &        1&         0&          0&     9354&        135&        1&       0&     48&    14.1&   -22.13&    13.92&   -21.69&      9.3&      170&    43.29    \nl
0517-3242     &        1&         1&          0&     3777&       50.9&        1&       0&       &   13.78&   -19.94&    13.37&   -20.15&         &         &             \nl
NGC 2110      &        1&         1&          1&     2284&       31.3&        2&      -3&     43&      14&         &    13.51&   -18.96&       17&      3.4&    42.30    \nl
MCG 8-11-11   &        0&         1&          0&     6141&       82.9&        1&       1&     28&      15&   -21.54&    14.62&   -19.95&       71&       29&    43.76    \nl
Mrk 3         &        1&         1&          1&     4050&       53.6&        2&      -2&     28&   14.03&   -20.24&    13.34&   -20.29&      347&       29&    44.07    \nl
NGC 2273      &        1&         1&          1&     1840&       28.4&        2&       1&     45&   12.55&   -20.25&    13.54&   -18.73&       33&      3.3&    42.50    \nl
Mrk 6         &        0&         1&          0&     5516&       78.4&        1&       0&     54&      15&   -19.97&    14.19&   -20.26&       75&       32&    43.73    \nl
0648+458      &        0&         1&          0&     6506&       88.7&        2&        &     67&    15.3&         &     15.3&   -19.41&         &         &             \nl
Mrk 376       &        0&         0&          0&    16770&      231&        1&        &       &   14.91&   -22.50&    14.62&   -22.14&      4.9&       16&    43.47    \nl
0714-2914     &        1&         1&          1&     1630&       24.1&        2&      -2&     56&   13.26&   -19.93&    13.52&   -18.38&         &         &             \nl
NGC 2377      &        1&         1&          1&     2457&       31.8&       L &       4&     41&   13.54&   -21.96&    14.77&   -17.74&     0.45&     0.69&    40.74    \nl
Mrk 78        &        0&         0&          0&    11202&      153&        2&       1&     55&      15&         &    14.58&   -21.30&       66&        5&    44.25    \nl
Mrk 79        &        1&         1&          0&     6652&       91.1&        1&       3&     47&    13.9&   -21.44&    14.27&   -20.50&       37&       55&    43.56    \nl
UGC 3995A     &        1&         1&          1&     4740&       66.1&        2&        &     53&    13.6&   -21.48&         &         &         &         &             \nl
Mrk 10        &        1&         0&          0&     8753&      119&        1&       3&     69&   13.34&   -22.47&    14.71&   -20.64&       14&       19&    43.37    \nl
{\bf Mrk 1210}&        1&         1&          1&     4046&       57.5&        2&        &      0&   14.34&   -19.57&       15&   -18.78&         &         &             \nl
Mrk 622       &        0&         1&          0&     6964&       96.2&        2&      -2&     26&    14.6&   -20.63&    14.08&   -20.81&        4&      0.3&    42.64    \nl
0816+2116     &        0&         1&          1&     5488&       76.9&        2&        &     59&   14.82&   -20.23&     15.5&   -18.91&         &         &             \nl
ESO 018-G009  &        0&         1&          0&     5341&       72.8&        2&        &       &        &         &         &         &         &         &             \nl
{\bf NGC 2639}&        1&         1&          1&     3336&       46.0&       L &       1&     54&   12.56&   -21.11&    11.88&   -21.42&     1.82&     0.48&    41.66    \nl
TON 951       &        1&         0&          0&    19200&      267&        1&        &       &      14&         &         &         &         &         &             \nl
NGC 2655      &        1&         1&          1&     1404&       24.4&       L &       0&     34&   10.96&   -21.12&    11.08&   -20.86&         &         &             \nl
NGC 2681      &        1&         1&          1&      692&       13.3&       U &       0&     24&   11.09&   -19.74&    11.33&   -19.29&         &         &             \nl
NGC 2691      &        1&         1&          1&     3981&       56.1&        1&       1&     50&   13.93&   -20.13&     14.1&   -19.63&         &         &             \nl
NGC 2768      &        1&         1&          1&     1335&       23.7&       L &      -5&     60&   10.84&   -21.17&     11.9&   -19.97&         &         &             \nl
NGC 2782      &        1&         1&          1&     2562&       37.3&        2&       1&     43&    12.3&   -20.85&    13.45&   -19.41&         &         &             \nl
NGC 2787      &        1&         1&          1&      696&       13.0&       U &      -1&     53&   11.82&   -18.96&    11.79&   -18.78&         &         &             \nl
Mrk 704       &        0&         0&          0&     8968&      125&        1&       1&     70&   15.38&   -20.57&     14.2&   -21.26&       13&       26&    43.37    \nl
0915+32       &        0&         0&          0&    18587&      259&       R &        &       &        &         &         &         &         &         &             \nl
NGC 2841      &        1&         1&          1&      638&       12.0&       L &       3&     67&   10.09&   -20.82&    12.41&   -17.99&     0.95&     0.63&    40.21    \nl
Mrk 705       &        0&         0&          0&     8658&      121&        1&      -2&     35&   14.88&   -20.89&     14.6&   -20.79&        9&       28&    43.19    \nl
NGC 2911      &        1&         1&          1&     3183&       46.8&       L &      -2&     40&    12.5&   -21.09&    13.83&   -19.51&         &         &             \nl
ESO 373-G013  &        0&         1&          0&     2700&       40.0&       L &        &       &        &         &         &         &         &         &             \nl
0936+368      &        0&         1&          0&     6021&       84.4&        1&        &     38&   15.35&   -19.42&         &         &         &         &             \nl
Mrk 403       &        0&         0&          0&     7225&      101&        2&        &     28&    15.4&         &    14.53&   -20.47&      9.4&     0.62&    43.05    \nl
\tablebreak
Mrk 1419      &        1&         1&          1&     4932&       70.6&       L &       1&     50&   13.29&   -21.22&       15&   -19.22&      4.4&         &    42.41    \nl
0939+2355     &        0&         1&          0&     6407&       90.2&        1&        &       &    15.3&         &         &         &         &         &             \nl
0942+09       &        0&         1&          1&     3897&       56.6&        2&        &       &    15.5&         &     15.5&   -18.25&         &         &             \nl
NGC 2992      &        1&         1&          1&     2314&       30.5&        2&       1&     62&   13.14&   -20.26&    13.78&   -18.64&       91&      8.1&    43.01    \nl
0945-30       &        1&         1&          0&     2482&       37.3&        2&       0&     65&   14.07&   -19.31&    13.69&   -19.16&     0.15&     0.37&    40.39    \nl
ESO 373-G29   &        1&         1&          0&     2802&       41.4&        2&       2&     59&      14&         &         &         &         &         &             \nl
NGC 2985      &        1&         1&          1&     1322&       22.4&       U &       2&     39&   11.18&   -20.77&     10.6&   -21.15&         &         &             \nl
NGC 3010B     &        0&         1&          1&     4760&       66.7&       L &        &     57&    15.1&         &         &         &         &         &             \nl
NGC 3032      &        0&         0&          0&     1533&       24.5&       O &      -2&     26&   13.18&   -19.11&         &         &         &         &             \nl
Mrk 1239      &        1&         1&          0&     5816&       82.8&        2&        &     38&    14.5&         &    14.39&   -20.18&       23&       17&    43.27    \nl
NGC 3031      &        1&         1&          1&      -34&        1.4&       L &       2&     60&    7.89&   -18.33&    11.63&   -14.10&     29.8&     6.38&    39.84    \nl
NGC 3081      &        1&         1&          1&     2394&       32.5&        2&       0&     33&   12.85&   -19.97&    13.55&   -19.01&      125&      8.1&    43.20    \nl
{\bf NGC 3079}      &        1&         1&          1&     1125&       20.4&       L &       7&     85&   11.54&   -21.14&    12.05&   -19.50&     0.18&         &    39.95    \nl
ESO 374-G025  &        0&         0&          0&     7006&       98.5&        1&        &       &   15.02&   -20.73&    15.29&   -19.65&         &         &             \nl
NGC 3125      &        0&         0&          0&      865&       18.6&       O &      -5&       &   13.26&   -18.24&    13.55&   -17.80&         &         &             \nl
NGC 3169      &        1&         1&          1&     1200&       19.7&       U &       1&       &   11.42&   -20.51&    12.41&   -19.06&         &         &             \nl
{\bf IC 2560} &        1&         1&          0&     2873&       42.7&        2&       3&     53&   12.53&   -21.22&    13.31&   -19.83&         &         &             \nl
NGC 3185      &        1&         1&          1&     1218&       21.3&       L &       1&     50&      12&   -18.99&         &         &      3.1&      0.6&    41.23    \nl
H1016+336     &        0&         0&          0&     7345&      103&        1&        &       &   15.93&         &    15.93&   -19.10&         &         &             \nl
NGC 3227      &        1&         1&          1&     1157&       20.6&        2&       1&     57&    11.1&   -20.39&    11.79&   -19.78&       64&       25&    42.51    \nl
ESO 568-G11   &        0&         0&          0&     8970&      126&       L &       4&     48&   14.56&         &         &         &         &         &             \nl
NGC 3281      &        1&         1&          0&     3460&       50.3&        2&       2&     59&    12.7&   -21.57&    14.02&   -19.47&      5.5&      0.5&    42.22    \nl
1034+0609     &        0&         1&          1&     3487&       51.4&        2&        &     50&    14.7&         &     14.7&   -18.84&         &         &             \nl
NGC 3312      &        1&         1&          1&     2869&       48.6&       L &       3&     70&   12.68&   -21.62&     12.6&   -20.83&        2&      0.7&    41.75    \nl
NGC 3362      &        1&         0&          0&     8318&      117&        2&       5&     39&   13.48&   -22.09&       15&   -20.31&      6.8&      0.7&    43.04    \nl
NGC 3367      &        1&         1&          1&     3038&       45.3&       U &       5&     29&   12.05&   -21.35&    14.14&   -19.13&         &         &             \nl
NGC 3393      &        1&         1&          1&     4107&       54.6&        2&       1&     25&   13.09&   -21.03&     13.1&   -20.57&         &         &             \nl
ESO 215-G014  &        0&         1&          0&     5581&       78.2&        1&        &       &        &         &     15.8&   -18.65&         &         &             \nl
NGC 3516      &        1&         1&          1&     2602&       38.9&        1&      -2&     44&    12.5&   -20.81&     12.4&   -20.55&       48&       73&    42.94    \nl
Fair 1150     &        1&         1&          0&     2930&       43.2&        1&       5&     32&   13.52&   -20.13&    13.11&   -20.07&         &         &             \nl
Mrk 732       &        1&         0&          0&     8768&      123&        1&      -4&      0&   14.04&   -21.53&       14&   -21.42&         &         &             \nl
NGC 3627      &        1&         1&          1&      727&        6.6&       U &       3&     65&    9.65&   -19.97&    11.94&   -17.16&         &         &             \nl
NGC 3642      &        1&         1&          1&     1588&       27.5&       L &       4&     34&   11.65&   -20.74&    14.04&   -18.16&     0.88&     0.51&    40.90    \nl
NGC 3660      &        0&         1&          1&     3678&       54.2&        2&       4&     36&    15.5&         &    14.45&   -19.20&         &         &             \nl
1121-2806     &        1&         1&          0&     4047&       58.7&        2&        &       &      13&         &         &         &         &         &             \nl
NGC 3665      &        0&         0&          0&     2080&       32.4&       R &      -2&     38&   11.77&   -20.86&         &         &         &         &             \nl
Mrk 40        &        0&         1&          1&     6323&       87.2&        1&      -2&     90&    16.8&         &    15.39&   -19.29&      7.2&      6.2&    42.81    \nl
ESO 439-G009  &        0&         0&          0&     7162&      105&        2&       1&       &   14.71&   -21.37&     14.8&   -20.28&         &         &             \nl
NGC 3718      &        1&         1&          1&     1247&       17.0&        1&       1&     63&   11.59&   -19.96&         &         &         &         &             \nl
NGC 3783      &        1&         1&          0&     3208&       43.3&        1&       1&     37&   12.64&   -21.13&    13.43&   -19.74&      130&      130&    43.46    \nl
NGC 3786      &        1&         1&          1&     2723&       41.6&        1&       1&     58&   13.24&   -20.13&    13.74&   -19.36&      8.4&        1&    42.24    \nl
Mrk 745       &        0&         1&          1&     3209&       47.4&        2&        &     50&    14.6&         &    14.62&   -18.75&         &         &             \nl
NGC 3884      &        1&         1&          0&     6948&       98.0&        1&       0&     53&    13.5&   -21.67&    12.88&   -22.05&         &         &             \nl
NGC 3898      &        1&         1&          1&     1176&       21.9&       L &       2&     55&    11.6&   -20.38&    11.83&   -19.87&         &         &             \nl
NGC 3921      &        1&         1&          1&     5838&       81.2&       L &       0&     53&   13.06&   -21.63&    13.35&   -21.17&     0.84&         &    41.81    \nl
Fair 1151     &        0&         0&          0&     3232&       47.3&       O &       1&       &   13.28&   -20.83&         &         &         &         &             \nl
Mrk 42        &        0&         0&          0&     7200&        100&        1&       3&     42&   15.28&   -19.95&    15.45&   -19.53&     1.22&      3.7&    42.15    \nl
NGC 3982      &        1&         1&          1&     1109&       17.0&        2&       3&     30&   11.78&   -19.47&     11.7&   -19.45&       15&      0.5&    41.71    \nl
NGC 3998      &        1&         1&          1&     1040&       21.6&       L &      -2&     36&   11.61&   -20.18&     12.1&   -19.57&     5.07&     2.39&    41.45    \nl
Mrk 1310      &        1&         1&          0&     5830&       83.3&        1&      -5&     53&   14.49&   -20.25&    15.46&   -19.12&      8.6&       11&    42.84    \nl
NGC 4036      &        1&         1&          1&     1397&       24.6&       L &      -3&     70&   11.57&   -20.46&     11.2&   -20.75&         &         &             \nl
Was 41        &        0&         0&          0&     7050&       99.4&        2&        &     30&      15&   -20.20&     14.9&   -20.06&         &         &             \nl
NGC 4051      &        1&         1&          1&      725&       17.0&        1&       4&     40&   10.83&   -20.41&    12.92&   -18.23&       39&       25&    42.13    \nl
NGC 4074      &        0&         1&          0&     6793&       95.9&        2&      -2&     54&    15.4&         &    14.44&   -20.44&       16&        1&    43.24    \nl
Was 45        &        1&         0&          0&     7494&      105&        2&        &      0&      14&   -20.97&    14.01&   -21.07&         &         &             \nl
NGC 4111      &        1&         1&          1&      807&       17.0&       L &      -1&     85&   11.63&   -19.55&    11.22&   -19.93&         &         &             \nl
NGC 4117      &        1&         1&          1&      871&       17.0&        2&      -2&     62&   14.04&   -17.12&     14.3&   -16.85&         &         &             \nl
Mrk 198       &        0&         0&          0&     7183&       99.8&        2&      -2&     31&   14.97&   -20.11&    14.73&   -20.24&     12.6&      1.5&    43.17    \nl
NGC 4151      &        1&         1&          1&      995&       20.3&        1&       2&     26&    11.5&   -20.83&    11.85&   -19.69&     1160&      442&    43.76    \nl
NGC 4192      &        1&         1&          1&     -142&       16.8&       L &       2&     78&   10.95&   -21.11&    13.26&   -17.87&         &         &             \nl
\tablebreak
Mrk 201       &        0&         0&          0&     2506&       39.1&       O &      10&     54&   13.01&   -20.10&     13.3&   -19.66&         &         &             \nl
Was 49B       &        0&         0&          0&    18900&      264&        1&        &       &        &         &     15.4&   -21.64&         &         &             \nl
NGC 4235      &        1&         1&          1&     2410&       35.1&        1&       1&     90&   12.62&   -20.84&     13.6&   -19.13&      2.4&      6.6&    41.55    \nl
NGC 4253      &        1&         1&          1&     3876&       55.1&        1&       1&     38&      14&   -19.90&    13.57&   -20.12&       59&       19&    43.32    \nl
{\bf NGC 4258}&        1&         1&          1&      448&        6.8&       L &       4&     70&     9.1&   -20.63&    11.65&   -17.51&      5.5&         &    40.48    \nl
NGC 4261      &        0&         0&          0&     2210&       35.1&       R &      -5&     29&   11.41&   -21.37&         &         &         &         &             \nl
NGC 4278      &        1&         1&          1&      649&        9.7&       L &      -5&     23&   11.09&   -18.96&    10.87&   -19.06&         &         &             \nl
NGC 4303      &        1&         1&          1&     1569&       15.2&       L &       4&     27&   10.18&   -20.79&     12.6&   -18.31&      3.1&      2.6&    40.93    \nl
Mrk 50        &        0&         0&          0&     7025&       99.4&        1&        &     39&      15&   -20.07&    15.17&   -19.79&         &         &             \nl
NGC 4388      &        1&         1&          1&     2524&       16.8&        2&       3&     78&   11.76&   -20.34&     13.9&   -17.23&       48&      3.8&    42.21    \nl
NGC 4395      &        1&         1&          1&      318&        3.6&        1&       9&     34&   10.64&   -17.21&     10.7&   -17.08&    12.59&     1.87&    40.29    \nl
NGC 4418      &        1&         1&          1&     2198&       33.7&       U &       1&       &   13.99&   -19.03&         &         &         &         &             \nl
NGC 4419      &        1&         1&          1&     -261&       16.8&       L &       1&     74&   12.08&   -19.53&    11.71&   -19.42&     0.29&         &    39.99    \nl
NGC 4438      &        1&         1&          1&       69&       16.8&       L &       0&     71&   11.02&   -20.64&    12.81&   -18.32&     0.97&     0.79&    40.52    \nl
NGC 4450      &        1&         1&          1&     1956&       16.8&       L &       2&     42&    10.9&   -20.38&    13.19&   -17.94&     0.53&         &    40.25    \nl
NGC 4486      &        1&         1&          1&     1282&       16.8&       L &      -4&     38&    9.59&   -21.64&    12.86&   -18.27&         &         &             \nl
NGC 4501      &        1&         1&          1&     2280&       16.8&       L &       3&     59&   10.36&   -21.27&    13.18&   -17.95&      2.9&      0.5&    40.99    \nl
NGC 4507      &        1&         1&          0&     3523&       50.7&        2&       2&     35&   12.92&   -21.23&    13.54&   -19.97&      110&     11.5&    43.52    \nl
NGC 4569      &        1&         1&          1&     -235&       16.8&       L &       2&     65&   10.26&   -21.34&    12.31&   -18.82&      5.4&      2.4&    41.26    \nl
NGC 4579      &        1&         1&          1&     1519&       16.8&       L &       3&     38&   10.48&   -20.84&    11.72&   -19.41&      1.1&     0.44&    40.57    \nl
NGC 4593      &        1&         1&          1&     2338&       39.5&        1&       3&     40&   11.67&   -21.55&    13.15&   -19.83&     17.2&       64&    42.51    \nl
NGC 4594      &        1&         1&          1&     1091&       20.0&        1&       1&     69&    8.98&   -23.13&     9.25&   -22.26&         &         &             \nl
Tol 74        &        1&         1&          0&     3285&       48.3&        2&       4&      0&      13&   -20.70&     13.5&   -19.91&         &         &             \nl
NGC 4639      &        1&         1&          1&     1010&       16.8&        1&       4&     49&   12.24&   -19.28&         &         &         &         &             \nl
1243+26       &        0&         0&          0&    26711&     372 &       R &        &       &    15.1&         &         &         &         &         &             \nl
ESO 172-G010  &        0&         1&          0&     1829&       24.7&        2&        &       &        &         &         &         &         &         &             \nl
NGC 4696      &        1&         1&          0&     2958&       46.4&       L &      -4&     46&   11.39&   -22.48&         &         &         &         &             \nl
1249-1308     &        0&         1&          0&     4385&       63.3&        1&        &       &      15&         &    14.03&   -19.96&       23&       26&    43.04    \nl
ESO 323-G32   &        1&         1&          0&     4857&       67.8&        2&      -1&     46&   13.86&   -20.84&     13.9&   -20.24&         &         &             \nl
NGC 4785      &        1&         1&          0&     3735&       53.4&        2&       3&     60&   13.21&   -21.83&     13.2&   -20.42&         &         &             \nl
Mrk 231       &        1&         0&          0&    12660&      169&        1&       5&     41&   14.41&   -22.12&    13.85&   -22.25&       23&       61&    43.88    \nl
NGC 4826      &        1&         1&          1&      408&        4.1&       L &       2&     59&    9.36&   -19.24&    12.14&   -15.92&     0.42&         &    38.93    \nl
1254+27       &        0&         0&          0&     7387&      103&       R &      -4&     64&   13.02&   -22.22&         &         &         &         &             \nl
1255-2930     &        1&         1&          0&     3055&       45.0&       L &       0&     51&   13.81&   -20.03&         &         &         &         &             \nl
NGC 4903      &        1&         1&          0&     4955&       70.6&       L &       5&     31&   13.72&   -21.09&     12.5&   -21.73&         &         &             \nl
NGC 4922B     &        0&         0&          0&     7153&       99.2&        2&      90&     56&    14.9&         &       15&   -19.96&      3.1&     0.85&    42.55    \nl
1300-3158     &        1&         1&          0&     4858&       69.2&       L &       3&     72&   13.95&   -21.19&         &         &         &         &             \nl
NGC 4941      &        1&         1&          1&     1111&        6.4&        2&       2&     60&    11.9&   -17.39&    12.23&   -16.80&       22&      1.3&    41.03    \nl
{\bf NGC 4945}&        1&         1&          1&      560&        5.2&        2&       6&     85&     9.3&   -21.15&     14.4&   -14.18&         &         &             \nl
1303-4008     &        1&         1&          0&     4477&       63.7&        1&      -2&     50&   13.56&   -20.91&    13.42&   -20.58&         &         &             \nl
ESO 508-G05   &        1&         1&          1&     2947&       43.5&        2&      -1&     41&   13.58&   -19.95&     13.6&   -19.58&         &         &             \nl
NGC 4968      &        1&         1&          0&     2957&       44.2&        2&      -2&     66&   13.75&   -19.81&     13.8&   -19.41&       10&      0.4&    42.36    \nl
1305-2407     &        0&         1&          0&     4227&       60.8&        2&        &       &      16&         &         &         &         &         &             \nl
NGC 5005      &        1&         1&          1&      946&       21.3&       L &       4&     63&   10.61&   -21.45&    13.67&   -17.97&     0.75&     0.13&    40.61    \nl
NGC 5033      &        1&         1&          1&      878&       18.7&        1&       5&     64&   10.75&   -21.15&    12.03&   -19.33&      5.3&     0.56&    41.35    \nl
IC 4218       &        0&         1&          0&     5815&       82.5&        1&        &     83&    14.6&   -20.93&    14.73&   -19.83&         &         &             \nl
1314-1532     &        0&         1&          1&     3900&       56.5&       U &        &       &      20&         &       20&   -13.74&         &         &             \nl
IC 4214       &        1&         1&          1&     2281&       34.3&       L &       1&     55&   12.16&   -21.10&         &         &         &         &             \nl
NGC 5077      &        1&         1&          1&     2832&       40.6&       L &      -5&     39&   12.38&   -20.83&    12.85&   -20.19&         &         &             \nl
NGC 5101      &        1&         1&          1&     1861&       27.4&       L &       0&     32&    11.6&   -21.18&         &         &         &         &             \nl
1319-1627     &        1&         1&          0&     5152&       71.9&        2&        &     76&    14.5&         &     13.9&   -20.36&      150&       13&    43.96    \nl
1321+31       &        0&         0&          0&     4782&       67.5&       R &        &     39&    12.9&   -21.32&         &         &         &         &             \nl
1322+2918     &        0&         0&          0&     7050&       98.5&        2&      -5&       &    16.2&         &         &         &         &         &             \nl
NGC 5128      &        0&         0&          0&      547&        4.9&       O &        &       &     7.7&   -21.15&    13.47&   -14.98&         &         &             \nl
NGC 5135      &        1&         1&          0&     4112&       58.8&        2&       2&     24&   12.88&   -21.46&    13.35&   -20.48&       22&      4.3&    42.95    \nl
NGC 5194      &        1&         1&          1&      463&        7.7&       L &       4&     54&    8.96&   -20.75&    13.47&   -15.96&      6.5&         &    40.66    \nl
NGC 5195      &        1&         1&          1&      465&        9.3&       L &      90&     38&   10.45&   -19.44&     12.3&   -17.54&         &         &             \nl
IC 894        &        0&         1&          0&     6507&       91.4&        2&        &     69&    14.9&         &         &         &         &         &             \nl
1331-2325     &        0&         1&          1&     2577&       39.3&        2&        &     65&   14.63&   -19.17&     14.7&   -18.26&      3.4&      1.1&    41.79    \nl
MCG -6-30-15  &        1&         1&          1&     2248&       35.1&        1&        &     55&    13.7&   -19.38&    13.61&   -19.10&     0.31&     0.02&    40.66    \nl
\tablebreak
\ 1335+39       &        1&         1&          0&     6023&       83.9&        2&        &      0&   14.27&   -20.42&     14.2&   -20.40&         &         &             \nl
NGC 5252      &        1&         1&          0&     6890&       96.9&        2&      -2&     61&   14.04&   -20.98&    14.21&   -20.70&         &         &             \nl
NGC 5256      &        1&         0&          0&     8353&      115&        2&      99&       &    14.1&         &    13.42&   -21.86&      3.2&     0.67&    42.69    \nl
NGC 5253      &        0&         0&          0&      404&        3.2&       O &      10&     71&   10.87&   -17.06&    13.16&   -14.37&         &         &             \nl
NGC 5283      &        1&         1&          1&     2700&       41.4&        2&      -2&     33&    14.2&   -18.97&    14.05&   -19.04&       27&      2.2&    42.74    \nl
NGC 5273      &        1&         1&          1&     1054&       21.3&        1&      -2&     33&   12.44&   -19.26&    13.12&   -18.52&      2.2&      0.2&    41.08    \nl
1344+038      &        0&         1&          0&     6870&       96.5&        2&        &       &    15.2&   -19.83&       16&   -18.90&         &         &             \nl
Mrk 461       &        0&         1&          1&     4856&       68.2&        2&        &     46&   14.61&   -19.80&     14.5&   -19.65&         &         &             \nl
NGC 5298      &        1&         1&          0&     4412&       62.8&       L &       3&     62&      14&   -20.71&         &         &         &         &             \nl
IC 4329A      &        1&         1&          0&     4813&       68.0&        1&      -1&     90&      14&   -20.47&    13.66&   -20.48&       34&       65&    43.27    \nl
{\bf NGC 5347}&        1&         1&          1&     2278&       36.7&        2&       2&     41&    13.4&   -19.72&     12.7&   -20.12&         &         &             \nl
NGC 5371      &        1&         1&          1&     2553&       37.8&       L &       4&     38&   11.32&   -21.72&    12.63&   -20.26&         &         &             \nl
Mrk 463       &        1&         0&          0&    14904&      211&        2&       1&     55&    14.4&   -22.53&    14.22&   -22.35&       73&       11&    44.57    \nl
NGC 5393      &        1&         1&          1&     5980&       84.0&       L &       1&     40&   13.97&   -21.02&         &         &         &         &             \nl
1357+28       &        0&         0&          0&    18850&      259&       R &        &       &        &         &         &         &         &         &             \nl
NGC 5427      &        1&         1&          1&     2730&       38.1&        2&       5&     32&   11.93&   -21.17&    13.96&   -18.94&      7.3&      0.9&    42.10    \nl
IC 4374       &        1&         1&          0&     6517&       91.3&       L &      -3&     42&   13.59&   -21.48&         &         &         &         &             \nl
Mrk 1370      &        0&         0&          0&     7370&      103&        2&        &       &      16&         &       16&   -19.04&         &         &             \nl
1408+1347     &        0&         1&          1&     4836&       68.3&        2&        &       &    15.2&         &         &         &         &         &             \nl
{\bf Circinus}&        1&         1&          1&      436&        4.2&        2&       3&     67&    12.1&   -19.62&         &         &      8.5&      0.6&    40.25    \nl
{\bf NGC 5506}&        1&         1&          1&     1815&       28.7&        2&       1&     90&   12.79&   -20.03&    14.38&   -17.91&       45&      5.5&    42.65    \nl
NGC 5548      &        1&         1&          1&     5149&       72.0&        1&       0&     33&    13.3&   -21.46&    13.73&   -20.54&       58&       83&    43.55    \nl
NGC 5635      &        1&         1&          1&     4316&       60.6&       L &        &     64&   13.46&   -20.97&         &         &         &         &             \nl
NGC 5643      &        1&         1&          1&     1199&       16.9&        2&       5&     23&   10.74&   -20.91&     13.6&   -17.54&       80&      4.8&    42.44    \nl
NGC 5675      &        1&         1&          1&     4169&       58.3&       L &        &     72&    13.7&   -20.87&       14&   -19.81&         &         &             \nl
NGC 5674      &        1&         0&          0&     7472&      104&        2&       5&     25&    13.7&   -21.49&     13.7&   -21.36&         &         &             \nl
Mrk 817       &        1&         0&          0&     9430&      128&        1&       0&     27&    14.5&   -21.15&    13.79&   -21.72&       13&       40&    43.40    \nl
NGC 5695      &        1&         1&          1&     4225&       58.9&        2&        &     44&   13.58&   -20.47&     13.6&   -20.24&      7.2&     0.61&    42.47    \nl
Mrk 477       &        0&         0&          0&    11340&      156&        2&        &     44&    15.4&   -20.70&    15.03&   -20.90&      150&       17&    44.62    \nl
NGC 5728      &        1&         1&          1&     2788&       42.2&        2&       1&     56&   12.37&   -21.41&     13.4&   -19.73&       68&      5.8&    43.16    \nl
ESO 512-G20   &        1&         1&          0&     3420&       48.7&        1&      -1&     47&   14.43&         &     14.5&   -18.93&         &         &             \nl
Mrk 1388      &        0&         1&          0&     6296&       87.3&        2&        &       &    15.7&         &       16&   -18.68&       26&      2.4&    43.37    \nl
Mrk 841       &        1&         0&          0&    10850&      150&        1&        &       &      14&         &    14.27&   -21.57&       25&       65&    43.81    \nl
1502+035      &        0&         1&          0&     5996&       83.4&        1&        &       &        &         &         &         &         &         &             \nl
NGC 5851      &        0&         1&          0&     6486&       89.9&       L &        &     77&   15.04&   -20.77&         &         &         &         &             \nl
NGC 5899      &        1&         1&          1&     2562&       39.8&       L &       5&     71&    12.5&   -21.22&    12.08&   -20.92&      6.9&      0.6&    42.12    \nl
NGC 5929      &        1&         1&          1&     2561&       38.5&        2&       2&     18&    14.1&         &     13.4&   -19.53&      9.3&      2.3&    42.22    \nl
1525+29       &        0&         0&          0&    19600&      270&       R &        &       &        &         &         &         &         &         &             \nl
1530-085      &        1&         1&          0&     6888&       96.4&        2&        &     70&    14.5&         &     15.7&   -19.19&         &         &             \nl
NGC 5953      &        1&         1&          1&     1965&       33.0&        2&       1&     37&    13.3&         &     13.1&   -19.49&      8.2&      2.7&    42.03    \nl
IC 4553       &        1&         1&          0&     5434&       74.9&        2&        &     38&   13.94&   -20.74&     14.4&   -19.95&         &         &             \nl
1533+14       &        0&         1&          0&     5910&       81.5&        2&        &     38&    15.3&   -19.42&     14.7&   -19.83&         &         &             \nl
Mrk 486       &        0&         0&          0&    11679&      159&        1&        &     57&    15.2&         &    14.78&   -21.19&      8.2&       36&    43.38    \nl
Mrk 860       &        0&         0&          0&     6869&       94.2&       O &        &     62&    15.5&         &     14.8&   -20.05&         &         &             \nl
NGC 5985      &        1&         1&          1&     2520&       39.2&       U &       3&     59&   11.87&   -21.59&         &         &         &         &             \nl
1545-1336     &        0&         1&          0&     2622&       37.0&        2&        &       &        &         &         &         &         &         &             \nl
1553+24       &        0&         0&          0&    12775&      175&       R &        &       &   15.42&   -21.14&         &         &         &         &             \nl
Mrk 298       &        0&         0&          0&    10098&      140&        2&      -1&     46&   15.17&   -20.75&    15.19&   -20.51&      4.9&      2.2&    43.05    \nl
1613+27       &        0&         0&          0&    19400&      267&       R &        &       &        &         &         &         &         &         &             \nl
ESO 0137-G34  &        1&         1&          1&     2620&       36.0&        2&       0&     39&   12.21&   -23.42&     12.2&   -20.57&         &         &             \nl
NGC 6217      &        1&         1&          1&     1362&       23.9&       U &       4&     34&   11.79&   -20.23&    13.88&   -18.01&         &         &             \nl
1637+27       &        0&         0&          0&    26400&      367&       R &        &       &        &         &         &         &         &         &             \nl
NGC 6251      &        1&         1&          0&     6900&       92.7&       L &      -5&     34&   13.64&   -21.59&         &         &         &         &             \nl
NGC 6211      &        1&         1&          0&     5966&       80.2&        2&      -1&     41&   13.63&         &     14.3&   -20.20&         &         &             \nl
ESO 0138-G01  &        1&         1&          1&     2740&       37.4&        2&        &     62&   14.31&   -19.52&    13.63&   -19.23&         &         &             \nl
NGC 6221      &        0&         0&          0&     1482&       19.4&       O &       5&     43&   10.66&   -21.67&    13.45&   -17.99&      6.2&      8.4&    41.45    \nl
NGC 6240      &        1&         0&          0&     7339&       98.9&        2&      90&     60&    13.8&         &    13.37&   -21.58&         &         &             \nl
NGC 6300      &        1&         1&          1&     1110&       14.3&        2&       3&     52&   10.98&   -20.58&    13.08&   -17.70&         &         &             \nl
Arp 102B      &        0&         0&          0&     7245&       97.4&        1&      -5&     40&   15.12&   -19.98&    14.81&   -20.11&         &         &             \nl
NGC 6328      &        1&         1&          0&     4260&       57.6&       L &       2&     56&   13.16&   -21.28&     15.5&   -18.29&         &         &             \nl
\tablebreak
ESO 139-G012  &        1&         1&          0&     5200&       70.2&        2&       4&       &   13.59&   -21.15&     13.6&   -20.61&         &         &             \nl
1736+32       &        0&         0&          0&    22200&      307&       R &        &       &        &         &         &         &         &         &             \nl
Ark 524       &        0&         0&          0&     7345&       98.5&        1&        &     38&    15.6&         &     15.6&   -19.34&         &         &             \nl
1752+32       &        0&         0&          0&    13460&      183&       R &        &       &        &         &         &         &         &         &             \nl
NGC 6500      &        1&         1&          1&     3003&       39.3&       L &       2&     44&   13.05&   -20.50&     13.7&   -19.26&         &         &             \nl
1827+32       &        0&         0&          0&    19770&      271&       R &        &       &    15.1&         &         &         &         &         &             \nl
Ark 539       &        0&         1&          1&     5065&       66.8&        2&        &     50&    15.4&   -19.03&       15&   -19.11&      1.3&     0.14&    41.83    \nl
1832-5926     &        1&         1&          0&     6065&       81.3&        2&        &       &    13.2&         &         &         &         &         &             \nl
{\bf ESO 103-G35}&        0&         1&          0&     3983&       53.2&        2&       0&     72&    14.7&   -19.30&    14.53&   -19.09&         &         &             \nl
IC 4729       &        1&         1&          0&     4419&       59.2&        2&        &       &   13.57&   -20.91&     11.8&   -22.04&         &         &             \nl
Fair 182      &        0&         1&          0&     3300&       43.9&        2&       0&     53&   14.99&   -18.65&     14.5&   -18.70&         &         &             \nl
Fair 51       &        0&         1&          0&     4210&       56.1&        1&       3&     61&    14.7&   -20.07&     14.1&   -19.63&       28&       32&    43.02    \nl
IC 4777       &        0&         1&          0&     5555&       74.0&        2&       0&     40&   14.59&   -20.01&    14.63&   -19.70&         &         &             \nl
3C 390.3      &        1&         0&          0&    16680&      228&        1&        &       &   14.37&         &    15.38&   -21.35&         &         &             \nl
Padova I      &        0&         1&          1&     2650&       33.5&        1&        &       &        &         &       15&   -17.62&         &         &             \nl
NGC 6764      &        1&         1&          1&     2416&       37.0&        2&       4&     57&   12.56&   -20.70&    14.03&   -18.81&      3.7&      6.2&    41.78    \nl
Fair 1163     &        0&         1&          0&     5640&       74.0&        2&       2&      0&   14.93&   -20.06&    14.97&   -19.36&         &         &             \nl
NGC 6814      &        1&         1&          1&     1563&       22.8&        1&       4&     20&   12.06&   -20.47&    14.21&   -17.58&     13.6&      4.4&    41.93    \nl
1953+0916     &        0&         1&          0&     7000&       91.3&        1&        &       &    15.1&         &         &         &         &         &             \nl
Cyg A         &        0&         0&          0&    16815&      228&        2&        &       &   17.04&   -22.15&     15.1&   -21.63&         &         &             \nl
NGC 6860      &        1&         1&          0&     4462&       58.8&        1&       3&     54&   13.68&   -20.62&    13.53&   -20.30&         &         &             \nl
20100-4156    &        0&         0&          0&    38850&      549&       L &        &       &        &         &         &         &         &         &             \nl
NGC 6890      &        1&         1&          1&     2419&       31.8&        2&       3&     39&   13.01&   -19.69&    14.02&   -18.49&       19&      0.9&    42.36    \nl
IC 4995       &        1&         1&          0&     4887&       61.5&        2&      -2&     50&   14.28&   -19.85&    14.06&   -19.87&         &         &             \nl
2031-307      &        0&         1&          0&     5700&       73.9&        1&        &       &        &         &         &         &         &         &             \nl
NGC 6951      &        1&         1&          1&     1426&       24.1&       L &       4&     36&   11.64&   -21.20&    14.04&   -17.87&     0.92&     0.05&    40.81    \nl
Mrk 509       &        1&         0&          0&    10200&      135&        1&      -3&     43&      13&         &    13.12&   -22.49&       81&      165&    44.23    \nl
IC 5063       &        1&         1&          1&     3402&       43.7&        2&      -1&     55&   12.89&   -20.61&     13.6&   -19.59&       93&        9&    43.32    \nl
NGC 7052      &        0&         0&          0&     4920&       62.1&       R &      -5&     58&    13.4&   -21.26&         &         &         &         &             \nl
IC 5135       &        1&         1&          0&     4842&       61.4&        2&       1&     29&   12.98&   -21.05&    13.87&   -20.06&         &         &             \nl
2157-3357     &        1&         1&          0&     4463&       56.4&       L &        &     83&    14.5&   -20.19&         &         &         &         &             \nl
2159-2244     &        0&         1&          0&     5313&       67.4&       L &       0&     62&   14.73&   -19.80&         &         &         &         &             \nl
NGC 7172      &        1&         1&          1&     2575&       33.9&        2&       1&     58&   12.85&   -20.11&    13.61&   -19.04&         &         &             \nl
ESO 108-G10   &        1&         1&          1&     3118&       40.8&        2&        &     55&   14.29&         &     15.8&   -17.24&         &         &             \nl
NGC 7212      &        0&         0&          0&     7795&        101&        2&       1&     60&   14.78&   -20.98&         &         &       70&      6.5&    43.92    \nl
NGC 7217      &        1&         1&          1&      946&       16.0&       L &       2&     36&   11.02&   -20.49&    11.92&   -19.10&     3.62&     2.02&    41.04    \nl
NGC 7213      &        1&         1&          0&     1792&       22.0&        1&       1&     26&   11.01&   -20.58&    12.08&   -19.63&         &         &             \nl
NGC 7214      &        1&         1&          0&     6832&       88.0&        1&       4&     52&   13.33&   -21.87&     14.1&   -20.60&      6.3&       11&    42.76    \nl
ESO 108-IG017 &        0&         1&          0&     1919&       24.7&        2&        &       &        &         &       14&   -17.96&         &         &             \nl
3C 449        &        0&         0&          0&     5122&       64.9&       O &      -3&       &   14.49&   -20.19&         &         &         &         &             \nl
NGC 7314      &        1&         1&          1&     1422&       18.3&        2&       4&     65&   11.62&   -20.22&    13.11&   -18.20&      6.1&     0.65&    41.39    \nl
NGC 7319      &        1&         1&          0&     6764&       85.0&        2&       4&     41&   14.11&   -21.02&    13.53&   -21.09&      6.4&     0.76&    42.73    \nl
Mrk 915       &        0&         0&          0&     7255&       93.2&        1&       1&     69&      15&         &     14.5&   -20.32&       46&       11&    43.67    \nl
2236+35       &        0&         0&          0&     8255&      107&       O &        &      0&   14.37&   -21.23&         &         &         &         &             \nl
2237+0747     &        1&         0&          0&     7375&       96.2&        1&       1&     30&   14.24&   -21.10&     14.3&   -20.59&         &         &             \nl
Mrk 917       &        0&         0&          0&     7317&       94.3&        2&       1&      0&    14.6&   -20.85&     14.1&   -20.75&       11&      1.3&    43.06    \nl
Ark 564       &        0&         0&          0&     7195&       95.4&        1&       3&     43&   14.55&   -21.00&    14.16&   -20.71&       19&       19&    43.31    \nl
NGC 7410      &        1&         1&          1&     1751&       20.1&        2&       1&     76&   11.24&   -20.72&     11.8&   -19.72&         &         &             \nl
NGC 7450      &        1&         1&          1&     3191&       38.1&        2&       1&     24&    14.5&         &    14.33&   -18.57&       12&      3.7&    42.32    \nl
Mrk 313       &        1&         1&          1&     1950&       21.5&        2&      -1&       &   13.31&   -18.58&       14&   -17.65&         &         &             \nl
NGC 7469      &        1&         1&          1&     4916&       60.9&        1&       1&     53&      13&   -21.27&    13.04&   -20.87&       58&       98&    43.40    \nl
NGC 7479      &        1&         1&          1&     2382&       32.4&       L &       5&     42&    11.6&   -21.33&    13.88&   -18.67&     1.09&     0.33&    41.14    \nl
NGC 7496      &        1&         1&          1&     1649&       20.1&        2&       3&     25&   11.91&   -19.68&     13.9&   -17.62&         &         &             \nl
NGC 7582      &        1&         1&          1&     1575&       17.6&        2&       2&     62&   11.37&   -20.40&    13.57&   -17.66&       35&       14&    42.11    \nl
NGC 7592      &        0&         0&          0&     7345&       93.9&       O &      -1&       &      15&         &    13.56&   -21.28&         &         &             \nl
NGC 7590      &        1&         1&          1&     1596&       17.3&        2&       4&     71&   12.11&   -19.73&    13.76&   -17.43&         &         &             \nl
{\bf IC 1481} &        1&         1&          1&     6118&       77.5&       L &        &     30&   14.39&   -20.42&         &         &         &         &             \nl
NGC 7672      &        0&         0&          0&     4117&       50.4&       O &       3&     42&    14.7&   -19.19&    14.21&   -19.29&         &         &             \nl
NGC 7674      &        1&         0&          0&     8698&      114&        2&       4&     23&   13.92&   -21.64&    14.36&   -20.89&       49&      3.8&    43.87    \nl
NGC 7678      &        1&         1&          1&     3500&       42.1&        2&       5&       &   12.66&   -20.94&     15.3&   -17.81&         &         &             \nl
\tablebreak
NGC 7682      &        1&         1&          0&     5120&       64.0&        2&       3&     36&    14.1&   -20.34&     14.3&   -19.71&       23&         &    43.05    \nl
IC 1495       &        1&         1&          0&     6375&       81.3&        2&       3&     41&    13.5&         &         &         &         &         &             \nl
NGC 7743      &        1&         1&          1&     1710&       24.4&       L &      -1&     31&   12.38&   -19.78&    13.28&   -18.66&     5.39&     0.67&    41.58    \nl
IC 1515       &        1&         1&          0&     6726&       85.9&        2&       2&     33&   14.45&   -20.48&         &         &         &         &             \nl

\tablecomments{This appendix lists some important parameters of the 364
galaxies observed for H$_2$O maser emission.  The columns list (1) the source
name, with galaxies detected as H$_2$O megamasers in boldface; 
(2 -- 4) indicators showing whether the galaxy is a member of the
magnitude-limited, distance-limited, or sensitivity-limited sample (a ``1''
is entered in the column if the galaxy is a member of the sample); (5) the
recession velocity in km s$^{-1}$ primarily from the NED database;
(6) the assumed luminosity 
distance in Mpc, derived from the recession velocity referred
to the rest frame of the 3\deg\ K background radiation (according to the
precepts in RC3) and based on H$_0$ = 75 km s$^{-1}$ Mpc$^{-1}$ and q$_0$ = 0,
except for the nearest galaxies, the distances of which were taken from 
Tully (1988); (7) the AGN classification taken primarily from
Huchra (1993) or V\'eron-Cetty \& V\'eron (1991),
where 1 = Seyfert 1, 2 = Seyfert 2, U = Unspecified Seyfert,
L = LINER, R = Radio galaxy, and O = Other type; (8) the Hubble type index, 
primarily from RC3; (9) galactic inclination in degrees, from Whittle (1992)
or derived from major and minor axis measurements (from NED) as described
in the text; (10) apparent B magnitude from RC3 (B$_T$ when available,
otherwise m$_B$), Huchra (1993) or NED;
(11) absolute B magnitude derived from apparent ``face-on'' magnitude (from
RC3); (12) apparent nuclear V magnitude from Huchra (1993)
or V\'eron-Cetty \& V\'eron (1991); (13) absolute nuclear V
magnitude; (14) [OIII]$\lambda$5007 line flux in 10$^{-14}$ erg s$^{-1}$
cm$^{-2}$ from Dahari \& De Robertis (1988), Ho, Filippenko \& Sargent (1993),
Stauffer (1982) or Whittle (1992);
(15) H$\beta$ line flux in 10$^{-14}$ erg s$^{-1}$ cm$^{-2}$
from Dahari \& De Robertis (1988), Ho, Filippenko \& Sargent (1993),
Stauffer (1982) or Whittle (1992);
(16) log of [OIII]$\lambda$5007 line luminosity in erg s$^{-1}$ derived from
column (14)
}
\end{planotable}

\pagestyle{empty}
\begin{planotable}{lrrrrrrrrrr}
\tablenum{1-b}
\tablecaption{Properties of Observed Galaxies}
\tablehead{
\colhead{Source} &
\colhead{S(12)} &
\colhead{S(25)} &
\colhead{S(60)} &
\colhead{S(100)} &
\colhead{P(12)} &
\colhead{P(25)} &
\colhead{P(60)} &
\colhead{P(100)} &
\colhead{S$_{6 cm}$} &
\colhead{log P$_{6 cm}$} \\
\colhead{} &
\colhead{(Jy)} &
\colhead{(Jy)} &
\colhead{(Jy)} &
\colhead{(Jy)} &
\colhead{(W Hz$^{-1}$)} &
\colhead{(W Hz$^{-1}$)} &
\colhead{(W Hz$^{-1}$)} &
\colhead{(W Hz$^{-1}$)} &
\colhead{(mJy)} &
\colhead{(W Hz$^{-1}$)} \\
\colhead{(1)} &
\colhead{(2)} &
\colhead{(3)} &
\colhead{(4)} &
\colhead{(5)} &
\colhead{(6)} &
\colhead{(7)} &
\colhead{(8)} &
\colhead{(9)} &
\colhead{(10)} &
\colhead{(11)} \\
}

\startdata

Mrk 334       &     0.23&      1.05&       4.35&     4.32&     23.28 &   23.94 &  24.56 & 24.56 &     8.5&    21.85\nl
Mrk 335       &     0.30&      0.38&       0.34& $<$0.57 &     23.54 &   23.64 &  23.60 & $<$23.&        &         \nl
Mrk 937       & $<$0.14 &  $<$0.24 &       0.52&     1.25&  $<$23.34 & $<$23.57&  23.91 & 24.29 &        &         \nl
Mrk 938       &     0.40&      2.37&      16.62&    17.16&     23.42 &   24.20 &  25.04 & 25.05 &        &         \nl
0017-5133     &         &          &           &         &           &         &        &       &        &         \nl
Mrk 945       &         &          &           &         &           &         &        &       &        &         \nl
NGC 224       &     1.70&      1.02&      10.79& $<$140.9&     20.00 &   19.78 &  20.80 & $<$21.&        &         \nl
NGC 235A      &         &          &           &         &           &         &        &       &        &         \nl
Mrk 348       &     0.31&      0.84&       1.29&     1.55&     23.06 &   23.50 &  23.69 & 23.77 &     480&    23.26\nl
Mrk 352       &         &          &           &         &           &         &        &       &        &         \nl
ESO 079-G16   &     0.18&      0.74&       2.98&     2.85&     23.11 &   23.73 &  24.34 & 24.32 &        &         \nl
NGC 404       & $<$0.11 &      0.20&       2.14&     3.98&  $<$19.10 &   20.15 &  21.17 & 21.44 &        &         \nl
Tol 0109-383  &     1.10&      1.74&       1.80&     1.82&     23.39 &   23.59 &  23.60 & 23.61 &    14.9&    21.52\nl
IC 1657       & $<$0.21 &      0.27&       2.77&     7.48&  $<$22.68 &   22.80 &  23.81 & 24.24 &        &         \nl
NGC 454       & $<$0.21 &  $<$0.42 &   $<$1.48 &     2.82&  $<$22.74 & $<$23.04& $<$23.5& 23.87 &        &         \nl
{\bf Mrk 1}         & $<$0.19 &      0.87&       2.53&     2.92&  $<$22.92 &   23.58 &  24.05 & 24.11 &      28&    22.09\nl
Mrk 565       & $<$0.21 &  $<$0.14 &       1.50&     3.42&  $<$23.07 & $<$22.89&  23.92 & 24.28 &        &         \nl
0120-0214     & $<$0.14 &      0.30&       1.69&     2.30&  $<$22.76 &   23.10 &  23.85 & 23.98 &        &         \nl
NGC 513       &     0.17&      0.28&       1.94&     4.05&     23.06 &   23.28 &  24.12 & 24.44 &        &         \nl
NGC 526A      &     0.28&      0.45&       0.22& $<$0.87 &     23.27 &   23.47 &  23.16 &       &       4&    21.42\nl
Mrk 993       & $<$0.13 &  $<$0.13 &       0.30&     1.32&  $<$22.73 & $<$22.72&  23.08 & 23.73 &        &         \nl
Mrk 359       &     0.12&      0.44&       1.13&     1.74&     22.76 &   23.32 &  23.73 & 23.92 &        &         \nl
0129-3323     &     0.18&      0.30&       2.16&     5.76&     22.94 &   23.15 &  24.01 & 24.44 &        &         \nl
NGC 600       & $<$0.12 &  $<$0.17 &       0.56&     2.57&  $<$21.88 & $<$22.02&  22.55 & 23.21 &        &         \nl
NGC 591       &     0.16&      0.45&       1.99&     3.48&     22.78 &   23.24 &  23.89 & 24.13 &     7.9&    21.49\nl
NGC 613       &     1.00&      2.45&      22.01&    49.95&     22.51 &   22.90 &  23.86 & 24.21 &        &         \nl
Mrk 1158      & $<$0.19 &      0.22&       0.97&     1.02&  $<$22.89 &   22.94 &  23.58 & 23.61 &        &         \nl
Mrk 573       & $<$0.25 &  $<$0.93 &       1.09&     1.26&  $<$23.09 & $<$23.66&  23.73 & 23.79 &     5.5&    21.43\nl
0147-0740     &     0.31&  $<$0.96 &       1.05&     0.65&     23.22 & $<$23.71&  23.75 & 23.54 &        &         \nl
0152+0622     & $<$0.08 &  $<$0.17 &       0.47&     1.20&  $<$22.59 & $<$22.92&  23.38 & 23.79 &        &         \nl
IC 184        & $<$0.10 &  $<$0.24 &       0.39&     0.91&  $<$22.75 & $<$23.12&  23.34 & 23.71 &        &         \nl
NGC 788       &     0.19&  $<$0.51 &       0.51&     0.59&     22.77 & $<$23.21&  23.21 & 23.27 &     1.2&    20.58\nl
ESO 153-IG16  & $<$0.06 &  $<$0.05 &       0.26& $<$0.98 &  $<$22.60 & $<$22.56&  23.26 & $<$23.&        &         \nl
ESO 153-G020  &     0.10&      0.20&       0.89& $<$3.02 &     22.84 &   23.15 &  23.81 & $<$24.&        &         \nl
NGC 824       &         &          &           &         &           &         &        &       &        &         \nl
4C 35.03      &         &          &           &         &           &         &        &       &        &         \nl
NGC 863       &     0.19&      0.22&       0.49&     1.46&     23.38 &   23.44 &  23.79 & 24.26 &     3.5&    21.64\nl
Ark 79        &         &          &           &         &           &         &        &       &        &         \nl
NGC 931       &     0.61&      1.32&       2.56&     4.55&     23.47 &   23.80 &  24.09 & 24.34 &        &         \nl
Mrk 1044      &     0.10&      0.22&       0.43&     0.88&     22.68 &   23.00 &  23.31 & 23.62 &        &         \nl
IC 1816       &     0.16&      0.39&       1.39&     2.34&     22.93 &   23.32 &  23.87 & 24.10 &        &         \nl
NGC 1019      & $<$0.14 &  $<$0.15 &       0.36& $<$1.37 &  $<$23.16 & $<$23.18&  23.57 & $<$24.&        &         \nl
{\bf NGC 1052}      &     0.20&      0.49&       0.90&     1.52&     21.89 &   22.27 &  22.53 & 22.76 &    1000&    22.58\nl
{\bf NGC 1068}      &    39.70&     85.04&     176.20&   224.00&     23.99 &   24.32 &  24.64 & 24.75 &    1090&    22.43\nl
NGC 1097      &     1.99&      5.51&      44.54&    85.34&     22.70 &   23.14 &  24.05 & 24.33 &        &         \nl
Mrk 372       & $<$0.13 &      0.17&       0.30& $<$0.19 &  $<$23.35 &   23.47 &  23.73 & $<$23.&        &         \nl
Mrk 1058      & $<$0.14 &      0.17&       0.59&     1.39&  $<$22.86 &   22.95 &  23.48 & 23.86 &        &         \nl
NGC 1125      &     0.17&      0.83&       3.31&     3.94&     22.55 &   23.23 &  23.83 & 23.91 &        &         \nl
NGC 1144      &     0.28&      0.63&       5.30&    11.34&     23.62 &   23.98 &  24.90 & 25.24 &        &         \nl
Mrk 1066      &     0.45&      2.26&      10.98&    12.15&     23.05 &   23.75 &  24.44 & 24.48 &    35.5&    21.95\nl
NGC 1167      &         &          &           &         &           &         &        &       &        &         \nl
H0258-015     &         &          &           &         &           &         &        &       &        &         \nl
NGC 1194      &     0.27&      0.51&       0.77& $<$0.93 &     22.90 &   23.19 &  23.37 & $<$23.&        &         \nl
NGC 1241      &     0.24&      0.45&       3.59&    10.32&     22.88 &   23.15 &  24.05 & 24.51 &        &         \nl
Mrk 1073      &     0.44&      1.41&       8.17&    11.11&     23.64 &   24.15 &  24.91 & 25.04 &        &         \nl
0314-2303     &     0.09&      0.15&       0.71&     1.96&     22.46 &   22.66 &  23.33 & 23.77 &        &         \nl
NGC 1275      &     1.07&      3.54&       7.15&     6.98&     23.77 &   24.29 &  24.60 & 24.59 &        &         \nl
NGC 1301      & $<$0.09 &      0.11&       0.52&     2.01&  $<$22.47 &   22.52 &  23.21 & 23.80 &        &         \nl
NGC 1326      &     0.32&      0.78&       8.06&    13.31&     22.03 &   22.43 &  23.44 & 23.66 &        &         \nl
Mrk 607       &     0.33&      1.07&       2.15&     2.75&     22.75 &   23.26 &  23.56 & 23.67 &        &         \nl
\tablebreak
MCG -2-9-40   & $<$0.12 &      0.33&       1.38&     3.53&  $<$22.67 &   23.11 &  23.74 & 24.15 &        &         \nl
Mrk 612       &     0.13&      0.27&       1.16&     1.81&     22.97 &   23.30 &  23.93 & 24.12 &        &         \nl
NGC 1358      & $<$0.08 &  $<$0.12 &       0.38&     0.93&  $<$22.42 & $<$22.59&  23.08 & 23.47 &     1.2&    20.58\nl
NGC 1365      &     3.37&     10.82&      76.13&   142.50&     23.06 &   23.57 &  24.42 & 24.69 &     191&    21.81\nl
{\bf NGC 1386}      &     0.49&      1.43&       5.40&     9.64&     22.23 &   22.69 &  23.27 & 23.52 &      13&    20.65\nl
III Zw 55     & $<$0.12 &      0.19&       0.87&     2.18&  $<$23.14 &   23.33 &  24.00 & 24.40 &        &         \nl
0339-2124     &     0.25&      0.10&       0.60&     1.49&     22.97 &   22.57 &  23.36 & 23.75 &        &         \nl
NGC 1433      &     0.24&      0.23&       3.49&    14.23&     21.58 &   21.57 &  22.75 & 23.36 &        &         \nl
NGC 1566      &     0.83&      1.22&      14.71&    46.36&     22.25 &   22.42 &  23.50 & 24.00 &        &         \nl
0425-0440     &     0.16&      1.41&       4.13&     3.30&     22.84 &   23.80 &  24.26 & 24.17 &        &         \nl
ESO 202-G23   &     0.08&      0.20&       1.87&     3.58&     22.63 &   23.01 &  23.98 & 24.27 &        &         \nl
NGC 1598      &     0.11&      0.19&       1.40&     4.22&     22.80 &   23.01 &  23.89 & 24.37 &        &         \nl
0438-0828     &     0.43&      1.62&       2.77&     2.53&     23.26 &   23.84 &  24.07 & 24.03 &        &         \nl
NGC 1672      &     1.67&      4.03&      32.96&    69.89&     22.62 &   23.01 &  23.92 & 24.25 &        &         \nl
NGC 1667      &     0.43&      0.68&       5.95&    14.73&     23.27 &   23.47 &  24.41 & 24.80 &       1&    20.64\nl
H0448-041     &         &          &           &         &           &         &        &       &        &         \nl
NGC 1685      & $<$0.10 &      0.22&       0.98&     1.53&  $<$22.62 &   22.97 &  23.62 & 23.82 &        &         \nl
0450-0317     &     0.12&      0.38&       0.83&     1.23&     22.76 &   23.25 &  23.59 & 23.76 &        &         \nl
0456+04       &     0.09&      0.27&       1.61&     3.20&     22.64 &   23.10 &  23.88 & 24.18 &        &         \nl
ESO 033-G02   &     0.21&      0.45&       0.68& $<$2.08 &     23.12 &   23.45 &  23.63 & $<$24.&        &         \nl
0459+0328     &         &          &           &         &           &         &        &       &        &         \nl
NGC 1808      &     4.44&     16.14&      87.81&   137.20&     22.79 &   23.35 &  24.09 & 24.28 &        &         \nl
UGC 3255      & $<$0.25 &  $<$0.25 &       1.14&     2.58&  $<$23.23 & $<$23.23&  23.89 & 24.25 &        &         \nl
ESO 362-G8    &     0.07&      0.18&       0.64& $<$1.50 &     22.51 &   22.95 &  23.50 & $<$23.&        &         \nl
H0510+031     &         &          &           &         &           &         &        &       &        &         \nl
Ark 120       &     0.32&      0.41&       0.64&     1.08&     23.83 &   23.93 &  24.13 & 24.36 &        &         \nl
0517-3242     &     0.22&      0.57&       1.40&     1.99&     22.84 &   23.24 &  23.63 & 23.79 &        &         \nl
NGC 2110      &     0.35&      0.84&       4.13&     5.68&     22.61 &   22.99 &  23.68 & 23.82 &     170&    22.30\nl
MCG 8-11-11   &     0.64&      1.95&       3.01&     4.24&     23.71 &   24.20 &  24.38 & 24.53 &      71&    22.76\nl
Mrk 3         &     0.71&      2.90&       3.77&     3.36&     23.38 &   23.99 &  24.11 & 24.06 &     361&    23.09\nl
NGC 2273      &     0.40&      1.36&       6.02&    10.00&     22.59 &   23.12 &  23.76 & 23.98 &      19&    21.26\nl
Mrk 6         &     0.22&      0.69&       1.18& $<$1.66 &     23.21 &   23.69 &  23.93 & $<$24.&     100&    22.86\nl
0648+458      & $<$0.17 &      0.44&       2.24&     3.41&  $<$23.21 &   23.61 &  24.31 & 24.50 &        &         \nl
Mrk 376       &     0.25&      0.51&       0.81&     1.39&     24.18 &   24.49 &  24.69 & 24.92 &        &         \nl
0714-2914     &     0.44&      1.77&       4.40&     6.04&     22.48 &   23.09 &  23.48 & 23.62 &    28.2&    21.29\nl
NGC 2377      &     0.42&      0.95&       8.93&    17.94&     22.71 &   23.06 &  24.03 & 24.34 &        &         \nl
Mrk 78        &     0.13&      0.56&       1.11&     1.13&     23.54 &   24.17 &  24.48 & 24.48 &      12&    22.51\nl
Mrk 79        &     0.31&      0.76&       1.50&     2.36&     23.47 &   23.87 &  24.16 & 24.36 &     4.2&    21.61\nl
UGC 3995A     &     0.22&      0.31&       0.56&     1.99&     23.06 &   23.20 &  23.46 & 24.01 &        &         \nl
Mrk 10        &     0.14&      0.22&       0.81&     1.95&     23.37 &   23.56 &  24.13 & 24.51 &        &         \nl
{\bf Mrk 1210}      &     0.50&      2.08&       1.89&     1.30&     23.29 &   23.91 &  23.87 & 23.71 &        &         \nl
Mrk 622       & $<$0.19 &      0.41&       1.28&     1.43&  $<$23.31 &   23.64 &  24.14 & 24.19 &        &         \nl
0816+2116     & $<$0.20 &      0.23&       0.81&     2.09&  $<$23.14 &   23.21 &  23.75 & 24.16 &        &         \nl
ESO 018-G009  &     0.17&      0.42&   $<$2.17 & $<$8.10 &     23.03 &   23.42 & $<$24.1& $<$24.&        &         \nl
{\bf NGC 2639}      &     0.16&      0.21&       1.99&     7.06&     22.61 &   22.72 &  23.70 & 24.25 &    54.5&    22.14\nl
TON 951       &         &          &           &         &           &         &        &       &        &         \nl
NGC 2655      &     0.18&      0.21&       1.67&     4.98&     22.10 &   22.18 &  23.08 & 23.55 &        &         \nl
NGC 2681      &     0.27&      0.56&       6.19&    11.77&     21.75 &   22.08 &  23.12 & 23.40 &        &         \nl
NGC 2691      &     0.17&      0.15&       1.37&     3.54&     22.79 &   22.73 &  23.71 & 24.12 & $<$0.6 & $<$20.34\nl
NGC 2768      &     0.09&  $<$0.08 &       0.37&     1.16&     21.78 & $<$21.71&  22.39 & 22.89 &        &         \nl
NGC 2782      &     0.49&      1.41&       8.67&    14.57&     22.92 &   23.37 &  24.16 & 24.38 &        &         \nl
NGC 2787      &     0.12&  $<$0.12 &       0.64& $<$3.01 &     21.38 & $<$21.38&  22.11 & $<$22.&        &         \nl
Mrk 704       &     0.35&      0.53&       0.36& $<$0.77 &     23.81 &   23.99 &  23.82 & $<$24.&        &         \nl
0915+32       &         &          &           &         &           &         &        &       &        &         \nl
NGC 2841      &     0.28&      0.22&       3.03&    18.77&     21.69 &   21.57 &  22.72 & 23.51 &        &         \nl
Mrk 705       &     0.15&      0.22&       0.59&     0.92&     23.42 &   23.56 &  24.00 & 24.20 &        &         \nl
NGC 2911      & $<$0.08 &  $<$0.24 &       0.22& $<$0.81 &  $<$22.32 & $<$22.80&  22.75 & $<$23.&        &         \nl
ESO 373-G013  &         &          &           &         &           &         &        &       &        &         \nl
0936+368      &         &          &           &         &           &         &        &       &        &         \nl
Mrk 403       &         &          &           &         &           &         &        &       &        &         \nl
\tablebreak
Mrk 1419      & $<$0.13 &  $<$0.26 &       0.71&     1.66&  $<$22.89 & $<$23.18&  23.62 & 23.99 &        &         \nl
0939+2355     & $<$0.15 &      0.33&       0.62&     0.72&  $<$23.17 &   23.50 &  23.77 & 23.84 &        &         \nl
0942+09       &         &          &           &         &           &         &        &       & $<$0.4 & $<$20.17\nl
NGC 2992      &     0.49&      1.26&   $<$8.50 & $<$29.36&     22.74 &   23.15 & $<$23.9& $<$24.&      77&    21.93\nl
0945-30       &         &          &           &         &           &         &        &       &       6&    21.00\nl
ESO 373-G29   &         &          &           &         &           &         &        &       &        &         \nl
NGC 2985      &     0.42&      0.46&       5.00&    19.94&     22.40 &   22.44 &  23.48 & 24.08 &        &         \nl
NGC 3010B     &         &          &           &         &           &         &        &       &        &         \nl
NGC 3032      & $<$0.25 &      0.19&       1.78&     4.26&  $<$22.26 &   22.13 &  23.11 & 23.49 &        &         \nl
Mrk 1239      &     0.65&      1.14&       1.23& $<$0.24 &     23.72 &   23.96 &  23.99 & $<$23.&        &         \nl
NGC 3031      &     0.75&      0.78&       6.81&    32.03&     20.24 &   20.26 &  21.20 & 21.88 &      58&    19.13\nl
NGC 3081      &         &          &           &         &           &         &        &       &     0.9&    20.06\nl
{\bf NGC 3079}      &     1.52&      2.27&      44.50&    89.22&     22.88 &   23.05 &  24.35 & 24.65 &      78&    21.59\nl
ESO 374-G025  &         &          &           &         &           &         &        &       &        &         \nl
NGC 3125      &     0.20&      0.79&       5.04&     5.13&     21.92 &   22.51 &  23.32 & 23.33 &        &         \nl
NGC 3169      &     0.59&      0.56&       6.51&    20.55&     22.44 &   22.41 &  23.48 & 23.98 &        &         \nl
{\bf IC 2560}       &     0.30&      0.94&       3.24&     6.11&     22.80 &   23.31 &  23.84 & 24.12 &        &         \nl
NGC 3185      &     0.16&      0.14&       1.43&     3.67&     21.93 &   21.88 &  22.89 & 23.30 &        &         \nl
H1016+336     &         &          &           &         &           &         &        &       &        &         \nl
NGC 3227      &     0.67&      1.76&       7.83&    17.59&     22.53 &   22.95 &  23.60 & 23.95 &      34&    21.24\nl
ESO 568-G11   & $<$0.07 &  $<$0.19 &       0.56&     1.32&  $<$23.12 & $<$23.53&  24.02 & 24.39 &        &         \nl
NGC 3281      &     0.89&      2.63&       6.86&     7.51&     23.42 &   23.90 &  24.31 & 24.35 &    26.7&    21.90\nl
1034+0609     & $<$0.16 &  $<$0.25 &       0.39& $<$0.11 &  $<$22.71 & $<$22.89&  23.08 & $<$22.& $<$0.4 & $<$20.09\nl
NGC 3312      & $<$0.18 &  $<$0.17 &       0.82&     3.36&  $<$22.71 & $<$22.68&  23.36 & 23.98 &        &         \nl
NGC 3362      &         &          &           &         &           &         &        &       &        &         \nl
NGC 3367      &     0.48&      1.09&       6.01&    12.56&     23.06 &   23.42 &  24.16 & 24.48 &        &         \nl
NGC 3393      &     0.13&      0.75&       2.25&     3.87&     22.66 &   23.42 &  23.90 & 24.13 &        &         \nl
ESO 215-G014  & $<$0.32 &  $<$0.25 &       0.34& $<$8.61 &  $<$23.36 & $<$23.25&  23.39 & $<$24.&        &         \nl
NGC 3516      &     0.43&      0.89&       1.76&     2.26&     22.89 &   23.21 &  23.50 & 23.61 &     4.3&    20.89\nl
Fair 1150     &     0.13&      0.24&       1.00&     2.93&     22.46 &   22.74 &  23.35 & 23.82 &        &         \nl
Mrk 732       &     0.16&  $<$0.25 &       1.75&     3.39&     23.45 & $<$23.65&  24.49 & 24.78 &        &         \nl
NGC 3627      &     4.96&      7.83&      67.80&   137.44&     22.41 &   22.61 &  23.55 & 23.86 &        &         \nl
NGC 3642      &     0.13&      0.10&       1.44&     4.58&     22.06 &   21.95 &  23.11 & 23.62 &        &         \nl
NGC 3660      &     0.19&      0.22&       1.87&     4.54&     22.83 &   22.89 &  23.81 & 24.20 &        &         \nl
1121-2806     &     0.16&      0.34&       0.57& $<$0.84 &     22.82 &   23.14 &  23.36 & $<$23.&        &         \nl
NGC 3665      & $<$0.25 &  $<$0.26 &       2.00&     6.37&  $<$22.50 & $<$22.51&  23.40 & 23.90 &        &         \nl
Mrk 40        &         &          &           &         &           &         &        &       &        &         \nl
ESO 439-G009  & $<$0.16 &      0.31&       0.65&     1.04&  $<$23.30 &   23.60 &  23.92 & 24.13 &        &         \nl
NGC 3718      &     0.17&      0.10&       0.70&     2.29&     21.77 &   21.53 &  22.38 & 22.90 &        &         \nl
NGC 3783      &     0.84&      2.49&       3.26&     4.90&     23.27 &   23.74 &  23.86 & 24.04 &      13&    21.46\nl
NGC 3786      &         &          &           &         &           &         &        &       &     3.4&    20.85\nl
Mrk 745       &         &          &           &         &           &         &        &       & $<$0.5 & $<$20.12\nl
NGC 3884      &         &          &           &         &           &         &        &       &        &         \nl
NGC 3898      &     0.11&  $<$0.07 &       0.42&     2.02&     21.80 & $<$21.63&  22.38 & 23.06 &        &         \nl
NGC 3921      & $<$0.12 &      0.11&       0.83&         &  $<$22.96 &   22.92 &  23.81 &       &        &         \nl
Fair 1151     &     0.53&      2.32&      32.68&    41.62&     23.15 &   23.79 &  24.94 & 25.04 &        &         \nl
Mrk 42        & $<$0.10 &  $<$0.14 &       0.32& $<$0.91 &  $<$23.06 & $<$23.21&  23.57 & $<$24.&        &         \nl
NGC 3982      &     0.51&      0.83&       6.57&    15.23&     22.24 &   22.46 &  23.36 & 23.72 &     2.2&    19.88\nl
NGC 3998      &     0.14&      0.16&       0.44&     0.93&     21.89 &   21.95 &  22.39 & 22.72 &        &         \nl
Mrk 1310      &         &          &           &         &           &         &        &       &        &         \nl
NGC 4036      &     0.09&  $<$0.10 &       0.50&     1.45&     21.79 & $<$21.85&  22.56 & 23.02 &        &         \nl
Was 41        &         &          &           &         &           &         &        &       &        &         \nl
NGC 4051      &     0.86&      1.59&       7.13&    23.92&     22.47 &   22.74 &  23.39 & 23.92 &       6&    20.32\nl
NGC 4074      &         &          &           &         &           &         &        &       &        &         \nl
Was 45        &     0.17&      0.38&       2.75&     5.57&     23.34 &   23.69 &  24.55 & 24.85 &        &         \nl
NGC 4111      &         &          &           &         &           &         &        &       &        &         \nl
NGC 4117      &         &          &           &         &           &         &        &       & $<$0.6 & $<$19.31\nl
Mrk 198       & $<$0.12 &      0.14&       0.62&     0.86&  $<$23.15 &   23.20 &  23.86 & 24.00 &        &         \nl
NGC 4151      &         &          &           &         &           &         &        &       &     125&    21.79\nl
NGC 4192      &     0.36&      0.44&       5.92&    19.68&     22.08 &   22.17 &  23.30 & 23.82 &        &         \nl
\tablebreak
Mrk 201       &     0.83&     46.77&      21.38&    25.88&     23.18 &   24.93 &  24.59 & 24.68 &      38&    21.84\nl
Was 49B       &         &          &           &         &           &         &        &       &        &         \nl
NGC 4235      &     0.00&      0.00&       0.32&     0.65&           &         &  22.67 & 22.98 &     5.3&    20.89\nl
NGC 4253      &     0.39&      1.30&       4.03&     4.66&     23.14 &   23.67 &  24.16 & 24.22 &    15.1&    21.73\nl
{\bf NGC 4258}      &     2.25&      2.81&      21.60&    78.39&     22.10 &   22.19 &  23.08 & 23.64 &     1.8&    19.00\nl
NGC 4261      &         &          &           &         &           &         &        &       &        &         \nl
NGC 4278      & $<$0.19 &  $<$0.08 &       0.56&     1.57&  $<$21.33 & $<$20.98&  21.80 & 22.25 &     290&    21.51\nl
NGC 4303      &     1.06&      1.40&      23.64&    64.65&     22.47 &   22.59 &  23.82 & 24.25 &        &         \nl
Mrk 50        &         &          &           &         &           &         &        &       &        &         \nl
NGC 4388      &     1.00&      3.46&      10.24&    18.10&     22.53 &   23.07 &  23.54 & 23.79 &      76&    21.41\nl
NGC 4395      & $<$0.13 &      0.21&       2.38&     6.34&  $<$20.30 &   20.50 &  21.57 & 21.99 &        &         \nl
NGC 4418      &     0.93&      9.32&      40.68&    32.80&     23.10 &   24.10 &  24.74 & 24.65 &        &         \nl
NGC 4419      &     0.54&      1.44&       7.99&    16.85&     22.26 &   22.69 &  23.43 & 23.76 &        &         \nl
NGC 4438      &     0.21&      0.17&       3.76&    11.27&     21.85 &   21.77 &  23.10 & 23.58 &        &         \nl
NGC 4450      &     0.11&  $<$0.13 &       1.34&     6.95&     21.57 & $<$21.64&  22.66 & 23.37 &        &         \nl
NGC 4486      &     0.23&  $<$0.24 &       0.39& $<$1.02 &     21.89 & $<$21.91&  22.12 & $<$22.&        &         \nl
NGC 4501      &     1.02&      1.28&      13.71&    54.65&     22.54 &   22.64 &  23.67 & 24.27 &        &         \nl
NGC 4507      &     0.46&      1.39&       4.31&     5.40&     23.14 &   23.63 &  24.12 & 24.22 &        &         \nl
NGC 4569      &     0.48&      1.05&       7.56&    23.66&     22.21 &   22.55 &  23.41 & 23.90 &        &         \nl
NGC 4579      &     0.35&      0.38&       4.74&    18.09&     22.07 &   22.11 &  23.20 & 23.79 &        &         \nl
NGC 4593      &     0.34&      0.81&       3.05&     5.95&     22.81 &   23.18 &  23.76 & 24.05 &     1.6&    20.48\nl
NGC 4594      &     0.39&      0.50&       3.11&    14.13&     22.27 &   22.38 &  23.17 & 23.83 &        &         \nl
Tol 74        &     0.64&      2.26&       7.52&    10.73&     23.25 &   23.80 &  24.32 & 24.47 & $>$13.6& $>$21.57\nl
NGC 4639      & $<$0.18 &  $<$0.22 &       1.41&     4.63&  $<$21.78 & $<$21.88&  22.68 & 23.19 &        &         \nl
1243+26       &         &          &           &         &           &         &        &       &        &         \nl
ESO 172-G010  & $<$0.26 &      0.26&       2.37&     3.99&  $<$22.28 &   22.28 &  23.24 & 23.46 &        &         \nl
NGC 4696      &         &          &           &         &           &         &        &       &        &         \nl
1249-1308     &     0.17&      0.37&       1.16&     2.22&     22.91 &   23.24 &  23.74 & 24.02 &        &         \nl
ESO 323-G32   &     0.11&      0.31&       0.99&     1.91&     22.77 &   23.22 &  23.73 & 24.01 &        &         \nl
NGC 4785      &     0.31&      0.48&       3.64&    10.47&     23.02 &   23.21 &  24.09 & 24.55 &        &         \nl
Mrk 231       &     1.87&      8.66&      31.99&    30.29&     24.79 &   25.46 &  26.02 & 26.00 &     170&    23.75\nl
NGC 4826      &     1.91&      2.54&      34.38&    74.50&     21.59 &   21.71 &  22.84 & 23.18 &        &         \nl
1254+27       &         &          &           &         &           &         &        &       &        &         \nl
1255-2930     &     0.24&      0.93&       5.96&     7.92&     22.76 &   23.35 &  24.15 & 24.28 &        &         \nl
NGC 4903      & $<$0.11 &      0.29&       0.77&     2.40&  $<$22.81 &   23.23 &  23.65 & 24.15 &        &         \nl
NGC 4922B     &     0.23&      1.29&       5.73&     7.54&     23.43 &   24.17 &  24.82 & 24.94 &        &         \nl
1300-3158     & $<$0.12 &  $<$0.15 &       0.58&     2.16&  $<$22.84 & $<$22.91&  23.51 & 24.09 &        &         \nl
NGC 4941      &     0.19&      0.53&       1.38&     4.19&     20.96 &   21.41 &  21.83 & 22.31 &     4.3&    19.32\nl
{\bf NGC 4945}      &     3.95&     14.45&     359.30&   620.50&     22.11 &   22.67 &  24.07 & 24.30 &        &         \nl
1303-4008     &     0.68&      1.22&       5.66&     8.79&     23.51 &   23.77 &  24.43 & 24.62 &        &         \nl
ESO 508-G05   & $<$0.13 &  $<$0.15 &       0.82&     2.31&  $<$22.46 & $<$22.53&  23.26 & 23.71 &        &         \nl
NGC 4968      &     0.39&      1.05&       2.38&     2.94&     22.96 &   23.38 &  23.74 & 23.83 &        &         \nl
1305-2407     & $<$0.15 &      0.67&       1.44&     1.49&  $<$22.80 &   23.47 &  23.80 & 23.81 &        &         \nl
NGC 5005      &     0.95&      1.21&      19.65&    54.26&     22.71 &   22.82 &  24.03 & 24.47 &        &         \nl
NGC 5033      &     0.95&      1.15&      13.80&    43.85&     22.60 &   22.68 &  23.76 & 24.26 &     3.3&    20.14\nl
IC 4218       &         &          &           &         &           &         &        &       &        &         \nl
1314-1532     &         &          &           &         &           &         &        &       &        &         \nl
IC 4214       &     0.27&      0.39&       4.28&    11.12&     22.58 &   22.73 &  23.78 & 24.19 &        &         \nl
NGC 5077      &         &          &           &         &           &         &        &       &        &         \nl
NGC 5101      &     0.12&  $<$0.17 &       0.89&     3.57&     22.01 & $<$22.18&  22.90 & 23.51 &        &         \nl
1319-1627     &     0.88&      2.86&       5.90&     5.48&     23.73 &   24.24 &  24.55 & 24.52 &        &         \nl
1321+31       &         &          &           &         &           &         &        &       &        &         \nl
1322+2918     &         &          &           &         &           &         &        &       &        &         \nl
NGC 5128      &    13.26&     17.26&     162.20&   313.80&     22.58 &   22.70 &  23.67 & 23.96 &        &         \nl
NGC 5135      &     0.64&      2.40&      16.91&    28.64&     23.42 &   23.99 &  24.84 & 25.07 &    58.8&    22.38\nl
NGC 5194      &     1.80&      2.61&      32.68&   137.70&     22.11 &   22.27 &  23.37 & 23.99 &    1.65&    19.07\nl
NGC 5195      &     0.72&      1.45&      10.35& $<$74.38&     21.87 &   22.18 &  23.03 & $<$23.&        &         \nl
IC 894        &         &          &           &         &           &         &        &       &        &         \nl
1331-2325     & $<$0.07 &  $<$0.26 &       0.83& $<$3.86 &  $<$22.13 & $<$22.67&  23.18 & $<$23.&        &         \nl
MCG -6-30-15  &     0.38&      0.81&       1.09&     1.10&     22.74 &   23.07 &  23.20 & 23.20 &       1&    20.16\nl
\tablebreak
\ 1335+39       &     0.12&      0.15&       0.97&     2.47&     23.00 &   23.08 &  23.90 & 24.31 &        &         \nl
NGC 5252      &         &          &           &         &           &         &        &       &        &         \nl
NGC 5256      &     0.23&      0.98&       7.34&    11.07&     23.55 &   24.18 &  25.05 & 25.23 &        &         \nl
NGC 5253      &     2.61&     11.96&      30.51&    29.36&     21.51 &   22.17 &  22.57 & 22.56 &        &         \nl
NGC 5283      &         &          &           &         &           &         &        &       &     5.7&    21.07\nl
NGC 5273      &         &          &           &         &           &         &        &       &     0.9&    19.69\nl
1344+038      &         &          &           &         &           &         &        &       &        &         \nl
Mrk 461       & $<$0.10 &      1.24&       3.75& $<$0.99 &  $<$22.73 &   23.83 &  24.31 & $<$23.&        &         \nl
NGC 5298      & $<$0.13 &  $<$0.09 &       0.43&     1.45&  $<$22.77 & $<$22.64&  23.30 & 23.83 &        &         \nl
IC 4329A      &     1.08&      2.21&       2.03&     1.66&     23.77 &   24.08 &  24.04 & 23.96 &      24&    22.12\nl
{\bf NGC 5347}      &     0.31&      0.96&       1.42&     2.64&     22.70 &   23.19 &  23.36 & 23.63 &     2.2&    20.55\nl
NGC 5371      &     0.21&      0.22&       3.31&    14.80&     22.55 &   22.58 &  23.75 & 24.40 &        &         \nl
Mrk 463       &     0.51&      1.58&       2.18&     1.92&     24.41 &   24.90 &  25.04 & 24.99 &     100&    23.71\nl
NGC 5393      & $<$0.15 &  $<$0.21 &       0.80&     2.44&  $<$23.09 & $<$23.23&  23.82 & 24.30 &        &         \nl
1357+28       &         &          &           &         &           &         &        &       &        &         \nl
NGC 5427      &     0.34&      0.59&   $<$5.49 & $<$24.19&     22.77 &   23.01 & $<$23.9& $<$24.&        &         \nl
IC 4374       &         &          &           &         &           &         &        &       &        &         \nl
Mrk 1370      &         &          &           &         &           &         &        &       &        &         \nl
1408+1347     &     0.13&      1.04&       3.70&     2.87&     22.86 &   23.76 &  24.31 & 24.20 &        &         \nl
{\bf Circinus}      &    18.80&     68.44&     248.70&   315.85&     22.60 &   23.16 &  23.72 & 23.82 &        &         \nl
{\bf NGC 5506}      &     1.28&      3.64&       8.41&     8.89&     23.10 &   23.55 &  23.92 & 23.94 &     160&    22.20\nl
NGC 5548      &     0.40&      0.77&       1.07&     1.61&     23.39 &   23.67 &  23.82 & 23.99 &        &         \nl
NGC 5635      &     0.08&  $<$0.11 &       0.29&     1.78&     22.56 & $<$22.69&  23.10 & 23.89 &        &         \nl
NGC 5643      &     1.10&      3.65&      19.49&    38.16&     22.57 &   23.10 &  23.82 & 24.12 &        &         \nl
NGC 5675      &     0.10&  $<$0.12 &       0.47&     1.81&     22.61 & $<$22.69&  23.28 & 23.86 &        &         \nl
NGC 5674      &     0.14&      0.28&       1.44&     3.73&     23.26 &   23.55 &  24.26 & 24.67 &        &         \nl
Mrk 817       &     0.34&      1.18&       2.12&     2.27&     23.81 &   24.35 &  24.61 & 24.64 &     5.3&    22.01\nl
NGC 5695      & $<$0.11 &      0.13&       0.57&     1.79&  $<$22.63 &   22.72 &  23.36 & 23.86 & $<$0.5 & $<$20.31\nl
Mrk 477       &     0.13&      0.51&       1.31&     1.85&     23.55 &   24.15 &  24.57 & 24.72 &      25&    22.85\nl
NGC 5728      &     0.24&      0.88&       8.16&    14.74&     22.71 &   23.27 &  24.24 & 24.50 &     4.6&    20.99\nl
ESO 512-G20   & $<$0.07 &  $<$0.18 &       0.28&     0.77&  $<$22.32 & $<$22.71&  22.89 & 23.34 &        &         \nl
Mrk 1388      & $<$0.15 &      0.23&       0.17& $<$0.57 &  $<$23.14 &   23.31 &  23.19 & $<$23.&        &         \nl
Mrk 841       &     0.19&      0.47&       0.46& $<$0.62 &     23.70 &   24.09 &  24.08 & $<$24.&        &         \nl
1502+035      &         &          &           &         &           &         &        &       &        &         \nl
NGC 5851      & $<$0.13 &      0.11&       0.62&     2.12&  $<$23.10 &   23.03 &  23.77 & 24.30 &        &         \nl
NGC 5899      &     0.26&      0.39&       3.41&    10.66&     22.69 &   22.86 &  23.81 & 24.31 &        &         \nl
NGC 5929      &     0.43&      1.62&       9.14&    13.69&     22.88 &   23.46 &  24.21 & 24.39 &    24.7&    21.64\nl
1525+29       &         &          &           &         &           &         &        &       &        &         \nl
1530-085      &     0.12&      0.36&       1.94&     3.00&     23.12 &   23.59 &  24.32 & 24.51 &        &         \nl
NGC 5953      &     0.53&      1.16&      10.04&    18.97&     22.84 &   23.18 &  24.12 & 24.39 &        &         \nl
IC 4553       &     0.48&      7.91&     103.80&   112.40&     23.50 &   24.72 &  25.84 & 25.87 &        &         \nl
1533+14       & $<$0.42 &  $<$0.13 &       0.26& $<$0.86 &  $<$23.51 & $<$22.99&  23.31 & $<$23.&        &         \nl
Mrk 486       &         &          &           &         &           &         &        &       &        &         \nl
Mrk 860       &     0.10&      0.54&       2.37&     2.68&     23.02 &   23.75 &  24.39 & 24.44 &        &         \nl
NGC 5985      &     0.13&      0.12&       1.05&     6.36&     22.39 &   22.33 &  23.29 & 24.07 &        &         \nl
1545-1336     &         &          &           &         &           &         &        &       &        &         \nl
1553+24       &         &          &           &         &           &         &        &       &        &         \nl
Mrk 298       &         &          &           &         &           &         &        &       &        &         \nl
1613+27       &         &          &           &         &           &         &        &       &        &         \nl
ESO 0137-G34  &     0.29&      0.74&       2.28& $<$20.62&     22.65 &   23.06 &  23.54 & $<$24.&        &         \nl
NGC 6217      &     0.51&      1.61&      10.83&    19.33&     22.54 &   23.04 &  23.87 & 24.12 &        &         \nl
1637+27       &         &          &           &         &           &         &        &       &        &         \nl
NGC 6251      &         &          &           &         &           &         &        &       &        &         \nl
NGC 6211      &         &          &           &         &           &         &        &       &        &         \nl
ESO 0138-G01  &     0.48&      1.76&       2.65& $<$2.39 &     22.90 &   23.47 &  23.64 & $<$23.&        &         \nl
NGC 6221      &     1.49&      5.27&      36.32&    84.50&     22.83 &   23.38 &  24.21 & 24.58 &        &         \nl
NGC 6240      &     0.56&      3.42&      22.68&    27.78&     23.80 &   24.59 &  25.41 & 25.50 &      80&    22.96\nl
NGC 6300      &     0.93&      2.27&      14.65&    36.03&     22.36 &   22.75 &  23.55 & 23.95 &        &         \nl
Arp 102B      &         &          &           &         &           &         &        &       &        &         \nl
NGC 6328      &         &          &           &         &           &         &        &       &        &         \nl
\tablebreak
ESO 139-G012  &     0.12&      0.17&       0.66&     1.61&     22.85 &   23.00 &  23.58 & 23.97 &        &         \nl
1736+32       &         &          &           &         &           &         &        &       &        &         \nl
Ark 524       &         &          &           &         &           &         &        &       &        &         \nl
1752+32       &         &          &           &         &           &         &        &       &        &         \nl
NGC 6500      &     0.10&      0.10&       0.64&     2.55&     22.27 &   22.27 &  23.07 & 23.67 &        &         \nl
1827+32       &         &          &           &         &           &         &        &       &        &         \nl
Ark 539       &         &          &           &         &           &         &        &       &        &         \nl
1832-5926     &     0.62&      1.39&       3.23&     3.91&     23.68 &   24.03 &  24.40 & 24.48 &        &         \nl
{\bf ESO 103-G35}   &     0.61&      2.36&       2.31&     1.05&     23.31 &   23.90 &  23.89 & 23.55 &        &         \nl
IC 4729       &     0.12&      0.16&       1.43&     4.49&     22.69 &   22.81 &  23.77 & 24.27 &        &         \nl
Fair 182      & $<$0.11 &      0.19&       0.93&     1.84&  $<$22.41 &   22.63 &  23.33 & 23.62 &        &         \nl
Fair 51       & $<$0.47 &      1.04&       1.84&     2.79&  $<$23.24 &   23.58 &  23.84 & 24.02 &        &         \nl
IC 4777       & $<$0.09 &  $<$0.14 &       0.61&     0.97&  $<$22.78 & $<$22.97&  23.60 & 23.79 &        &         \nl
3C 390.3      &     0.13&      0.29&       0.20& $<$0.62 &     23.88 &   24.23 &  24.08 & $<$24.&        &         \nl
Padova I      &         &          &           &         &           &         &        &       &        &         \nl
NGC 6764      &     0.36&      1.29&       6.33&    11.56&     22.77 &   23.33 &  24.02 & 24.28 &      35&    21.76\nl
Fair 1163     &         &          &           &         &           &         &        &       &        &         \nl
NGC 6814      & $<$0.49 &      0.60&       5.52&    18.88&  $<$22.48 &   22.57 &  23.54 & 24.07 &     2.2&    20.14\nl
1953+0916     & $<$0.66 &  $<$0.25 &       0.52& $<$7.38 &  $<$23.81 & $<$23.39&  23.71 & $<$24.&        &         \nl
Cyg A         & $<$0.25 &      1.06&       2.33& $<$8.28 &  $<$24.17 &   24.80 &  25.14 & $<$25.&        &         \nl
NGC 6860      &     0.24&      0.33&       0.95&     2.47&     22.99 &   23.13 &  23.59 & 24.00 &        &         \nl
20100-4156    &         &          &           &         &           &         &        &       &        &         \nl
NGC 6890      &     0.34&      0.65&       3.86&     8.16&     22.62 &   22.90 &  23.67 & 23.99 &     4.2&    20.71\nl
IC 4995       &     0.09&      0.33&       0.83&     1.25&     22.58 &   23.17 &  23.57 & 23.75 &        &         \nl
2031-307      &         &          &           &         &           &         &        &       &        &         \nl
NGC 6951      &     0.62&      1.37&      13.21&    37.47&     22.63 &   22.98 &  23.96 & 24.42 &        &         \nl
Mrk 509       &     0.32&      0.70&       1.36&     1.52&     23.82 &   24.17 &  24.46 & 24.50 &     4.5&    21.98\nl
IC 5063       &     1.07&      3.91&       5.34&     4.17&     23.38 &   23.95 &  24.08 & 23.97 &        &         \nl
NGC 7052      & $<$0.25 &  $<$0.25 &       0.45&     1.48&  $<$23.06 & $<$23.06&  23.31 & 23.83 &        &         \nl
IC 5135       &     0.59&      2.12&      16.48&    25.57&     23.42 &   23.97 &  24.87 & 25.06 &        &         \nl
2157-3357     & $<$0.12 &  $<$0.16 &       0.42&     1.05&  $<$22.65 & $<$22.78&  23.20 & 23.60 &        &         \nl
2159-2244     & $<$0.12 &  $<$0.18 &       0.25&     0.70&  $<$22.80 & $<$22.97&  23.12 & 23.57 &        &         \nl
NGC 7172      &     0.44&      0.76&       5.71&    12.29&     22.78 &   23.02 &  23.90 & 24.23 &     1.8&    20.39\nl
ESO 108-G10   & $<$0.10 &      0.13&       1.00&     1.64&  $<$22.31 &   22.42 &  23.30 & 23.51 &        &         \nl
NGC 7212      &     0.20&      0.77&       2.89&     4.90&     23.36 &   23.96 &  24.53 & 24.76 &        &         \nl
NGC 7217      &     0.38&      0.30&       4.96&    18.45&     22.06 &   21.96 &  23.18 & 23.75 &        &         \nl
NGC 7213      &     0.61&      0.74&       2.67&     8.18&     22.55 &   22.63 &  23.19 & 23.68 &        &         \nl
NGC 7214      &     0.16&      0.37&       1.92&     5.33&     23.16 &   23.53 &  24.24 & 24.68 &        &         \nl
ESO 108-IG017 & $<$0.09 &      0.20&       1.37&     2.17&  $<$21.80 &   22.16 &  23.00 & 23.20 &        &         \nl
3C 449        &         &          &           &         &           &         &        &       &        &         \nl
NGC 7314      &     0.27&      0.58&       3.74&    14.15&     22.03 &   22.37 &  23.18 & 23.75 &     1.9&    19.88\nl
NGC 7319      & $<$0.16 &  $<$0.23 &       0.88&     2.54&  $<$23.12 & $<$23.29&  23.87 & 24.33 &        &         \nl
Mrk 915       & $<$0.17 &      0.35&       0.46&     1.01&  $<$23.25 &   23.56 &  23.67 & 24.01 &        &         \nl
2236+35       &         &          &           &         &           &         &        &       &        &         \nl
2237+0747     &     0.20&      0.34&       0.83& $<$2.48 &     23.34 &   23.57 &  23.95 & $<$24.&        &         \nl
Mrk 917       &     0.19&      0.61&       3.71&     5.95&     23.29 &   23.80 &  24.59 & 24.79 &        &         \nl
Ark 564       & $<$0.30 &      0.57&       0.83&     1.14&  $<$23.50 &   23.78 &  23.94 & 24.08 &      11&    22.07\nl
NGC 7410      &     0.11&  $<$0.08 &       0.71&     2.97&     21.71 & $<$21.61&  22.54 & 23.16 &        &         \nl
NGC 7450      &         &          &           &         &           &         &        &       &     1.7&    20.47\nl
Mrk 313       &     0.28&      0.39&   $<$4.94 & $<$13.70&     22.19 &   22.33 & $<$23.4& $<$23.&        &         \nl
NGC 7469      &     1.35&      5.79&      25.87&    34.90&     23.77 &   24.40 &  25.05 & 25.18 &      65&    22.45\nl
NGC 7479      &     0.88&      3.68&      12.84&    23.56&     23.04 &   23.67 &  24.21 & 24.47 &        &         \nl
NGC 7496      &     0.35&      1.60&       8.46&    15.55&     22.22 &   22.89 &  23.61 & 23.88 &        &         \nl
NGC 7582      &     1.62&      6.44&      49.10&    72.92&     22.78 &   23.38 &  24.26 & 24.43 &      69&    21.41\nl
NGC 7592      &     0.28&      0.90&       7.80&    10.25&     23.46 &   23.97 &  24.91 & 25.02 &        &         \nl
NGC 7590      &     0.56&      0.83&       6.91& $<$23.70&     22.30 &   22.47 &  23.39 & $<$23.&        &         \nl
{\bf IC 1481}       & $<$0.13 &      0.28&       1.41&     1.51&  $<$22.98 &   23.29 &  24.00 & 24.03 &        &         \nl
NGC 7672      & $<$0.09 &  $<$0.15 &       0.46& $<$2.46 &  $<$22.44 & $<$22.67&  23.14 & $<$23.&       1&    20.48\nl
NGC 7674      &     0.67&      1.90&       5.59&     8.15&     24.00 &   24.45 &  24.92 & 25.09 &        &         \nl
NGC 7678      &     0.41&      0.87&       6.59&    14.69&     22.93 &   23.26 &  24.14 & 24.49 &        &         \nl
\tablebreak
NGC 7682      &         &          &           &         &           &         &        &       &        &         \nl
IC 1495       & $<$0.12 &  $<$0.36 &       0.94& $<$4.27 &  $<$22.97 & $<$23.44&  23.86 & $<$24.&        &         \nl
NGC 7743      & $<$0.11 &  $<$0.20 &       0.79&     2.53&  $<$21.90 & $<$22.15&  22.75 & 23.26 &        &         \nl
IC 1515       & $<$0.15 &  $<$0.23 &       0.57&     1.33&  $<$23.12 & $<$23.31&  23.69 & 24.06 &        &         \nl

\tablecomments{This appendix lists some important parameters of the 364
galaxies observed for H$_2$O maser emission.  The columns list (1) the source
name, with galaxies detected as H$_2$O megamasers in boldface; 
(2 -- 5) the IRAS flux at 12, 25, 60 and 100 $\mu m$, in Jy, from
the IRAS Faint Source Catalog (e.g. Moshir et al. 1989);
(6 -- 9) the log of power derived from the IRAS fluxes at 12, 25, 60 and 100
$\mu m$, in W Hz$^{-1}$; (10) the 6 cm nuclear radio continuum flux in mJy
from Unger et al. 1987, Ulvestad \& Wilson 1984a, 1984b, or 1989;
(11) the log of power, in W Hz$^{-1}$, derived from the 6 cm nuclear radio
continuum flux.
}

\end{planotable}

\pagestyle{empty}
\begin{planotable}{lrrrrrccccccc}
\tablenum{3}
%\tablewidth{650pt}
\tablecaption{Results of Statistical Comparisons}
\tablehead{
\colhead{Prop} &
\colhead{Fig} &
\colhead{Det} &
\colhead{M} &
\colhead{D} &
\colhead{S} &
\colhead{Cens} &
\colhead{K-S} &
\colhead{Gehan} &
\colhead{Gehan} &
\colhead{Logrank} &
\colhead{P-P} &
\colhead{P-Pr} \\
 & & \colhead{Gal's} & & & & \colhead{Entries} & & \colhead{A} & \colhead{B} &
 & & \\
\colhead{(1)} &
\colhead{(2)} &
\colhead{(3)} &
\colhead{(4)} &
\colhead{(5)} &
\colhead{(6)} &
\colhead{(7)} &
\colhead{(8)} &
\colhead{(9)} &
\colhead{(10)} &
\colhead{(11)} &
\colhead{(12)} &
\colhead{(13)} \\
}

\startdata

Recession Velocity & 2    &15  &228  &263  &150 & - & .09 & .01 & $<$.01 &
.01 & - & -     \nl
Recession Velocity\tablenotemark{a} & 2    & 15 & 228 & 263 &150& -  &
.65 & .42 & .39 & .72 & - & -     \nl
m$_B$         & 6    &15  &228  &252  &149 & - & .10 & .10 & .07 &
.14 & - & -     \nl
m$_B$\tablenotemark{a} & 6    & 15 & 228 & 252 & 149& - & .84 &
.79 & .78 & .81 & - & -     \nl
m$_B$\tablenotemark{b}& - &15  & - &187 & - & - & .07 & .09 & .06 &
.13 & - & -     \nl
M$_B$         & 7    &15  &192  &201  &128 & 51    & - & .12 & .11 &
.11 & .12 & .13 \nl
M$_B$     \tablenotemark{b}&-&15 &- &187  &- & 38    & - &.13 & .12 &
.13 & .13 & .14 \nl
Hubble T      & 8   &13  &203  &207  &128 & -     & - & .47 & .49 &
.58 & - & -     \nl
Inclination   & 9   &14  &192  &210  &132 & - & .12 & .19 & .21 &
.06 & - & -     \nl
Inclination\tablenotemark{b}&-&14  &-&165  &-& - & .13 & .18 & .20 &
.03 & - & -     \nl
IRAS S(12)    & 13   &13  &165  &158  &101 & 58 & -    & $<$.01 & $<$.01 &
$<$.01 & $<$.01 & $<$.01 \nl
IRAS P(12)    & 14   &13  &165  &158  &101 & 58 & -    & .51 & .50 &
.28 & .48 & .49 \nl
IRAS S(25)    & 15   &15  &163  &170  &100 & 44 & -    & $<$.01 & $<$.01 &
$<$.01 & $<$.01 & $<$.01 \nl
IRAS P(25)    & 16   &15  &163  &170  &100 & 44 & -    & .13 & .11 &
.05 & .12 & .13 \nl
IRAS S(60)    & 17   &15  &195  &210  &124 & 5 & -     & .01 & $<$.01 &
$<$.01 & .01 & .01 \nl
IRAS P(60)    & 18   &15  &195  &210  &124 & 5 & -     & .25 & .23 &
.25 & .25 & .25     \nl
IRAS S(100)   & 19   &15  &177  &186  &113 & 28 & -    & .12 & .09 &
.18 & .13 & .10 \nl
IRAS P(100)   & 20   &15  &177  &186  &113 & 28 & -    & .98 & .98 &
.70 & $>$.99 & $>$.99 \nl
IRAS S(12)/S(25)& 22 &13  & -   &146  & - & - & .34 & .22 & .19 &
.13 & - & -   \nl
IRAS S(60)/S(100)& 23&15  &176  &185  &113 & - & .02 & .01 & .03 &
.01 & - & -  \nl
IRAS S(60)/S(100)\tablenotemark{b}& 24&15  & -   &140  & -  & - & .02 &
.02 & .03 & .02 & - & -  \nl
IRAS S(25)/S(60)& 25 &15  &160  &166  &97 & - & .13 & .16 & .18 &
.12 & - & -   \nl
Nuclear m$_V$ & 26   &13  &198  &218  &137 & - & .17 & .26 & .21 &
.47 & - & -     \nl
Nuclear M$_V$  & 27   &13  &198  &218  &137& -  & .44 & .25 & .28 &
.29 & - & -     \nl
S(6 cm)       & 28   &9  &56   &54   &42  & 6 & -     & .04 & .02 &
.03 & .04 & .03 \nl
P(6 cm)       & 29   &9  &56   &54   &42  & 6 & -     & .12 & .08 &
.51 & .13 & .10 \nl
S([OIII])     & 30   & 8  &112  &113  &81  & - & .59 & .48 & .49 &
.18 & - & -     \nl
P([OIII])     & 31   & 8  &112  &113  &81  & - & .52 & .39 & .35 &
.65 & - & -     \nl
Total sample  &  -     &15  &228  &263  &150 & - &-&-&-&-&-&-\nl
\tablenotetext{a}{The significance values in these rows describe a comparison
between the galaxies detected in H$_2$O maser emission and the undetected
galaxies in the sensitivity-limited sample.  The values in all
other rows describe a comparison of the detected galaxies with the
undetected galaxies in the distance-limited sample.}
\tablenotetext{b}{In these rows, the H$_2$O-detected galaxies are compared to
the undetected galaxies from the distance-limited sample which are classified
as either a Seyfert 2 or LINER (i.e. Seyfert 1s are omitted).}
\tablecomments{The columns list (1) the physical property being compared, (2)
the corresponding figure showing a histogram of the comparison, (3) the
number of detected H$_2$O maser galaxies included in the comparison,
(4) the number of undetected galaxies from the magnitude-limited sample
represented in the histogram,
(5) the number of undetected galaxies
from the distance-limited sample represented in the histogram,
(6) the number of undetected galaxies from the 
sensitivity-limited sample represented in the histogram,
(7) the number of censored data values (i.e. only a limit, rather than a 
detection, is available) used in the statistical calculations
(8) the probability that one parent population would produce the
detected and undetected galaxy samples, based on the Kolmogorov-Smirnov test,
(9) the probability that one parent population would produce the
detected and undetected galaxy samples, based on Gehan's generalized Wilcoxon
test with permutation variance,
(10) the probability that one parent population would produce the
detected and undetected galaxy samples, based on Gehan's generalized Wilcoxon 
test with hypergeometric variance,
(11) the probability that one parent population would produce the
detected and undetected galaxy samples, based on the logrank test,
(12) the probability that one parent population would produce the
detected and undetected galaxy samples, based on the Peto-Peto test,
(13) the probability that one parent population would produce the
detected and undetected galaxy samples, based on the Peto-Prentice test.
}
\end{planotable}